\documentclass[journal,10pt]{IEEEtran}
\usepackage{graphicx}  
\usepackage{stfloats}
\usepackage{subfigure}
\usepackage{array}
\usepackage{fontenc}
\usepackage{balance}
\usepackage{amsmath,amssymb,amsthm,amsfonts}
\usepackage{mathtools}
\usepackage{url}
\usepackage{multicol}
\usepackage{multirow}
\usepackage{bm}
\usepackage{upgreek}
\usepackage{textcomp}
\usepackage{algpseudocode,algorithm}

\begin{document}
\title{Sparsity Based Autoencoders for Denoising Cluttered Radar Signatures}
\title{Optimization of Radar Parameters for Maximum Detection Probability Under Generalized Discrete Clutter Conditions Using Stochastic Geometry}
\author{Shobha~Sundar~Ram, Shelly~Vishwakarma, Akanksha~Sneh and Kainat~Yasmeen,
\thanks{Authors are with the Indraprastha Institute of Information Technology Delhi, New Delhi 110020 India. E-mail: \{shobha@iiitd.ac.in}%
}
\maketitle

\begin{abstract}
Narrowband and broadband indoor radar images significantly deteriorate in the presence of target dependent and independent static and dynamic clutter arising from walls. 
A stacked and sparse denoising autoencoder (StackedSDAE) is proposed for mitigating wall clutter in indoor radar images. The algorithm relies on the availability of clean images and corresponding noisy images during training and requires no additional information regarding the wall characteristics. The algorithm is evaluated on simulated Doppler-time spectrograms and high range resolution profiles generated for diverse radar frequencies and wall characteristics in around-the-corner radar (ACR) scenarios. Additional experiments are performed on range-enhanced frontal images generated from measurements gathered from a wideband RF imaging sensor. The results from the experiments show that the StackedSDAE successfully reconstructs images that closely resemble those that would be obtained in free space conditions. Further, the incorporation of sparsity and depth in the hidden layer representations within the autoencoder makes the algorithm more robust to low signal to noise ratio (SNR) and label mismatch between clean and corrupt data during training than the conventional single layer DAE. For example, the denoised ACR signatures show a structural similarity above 0.75 to clean free space images at SNR of $-10dB$ and label mismatch error of 50\%.
\end{abstract}
\maketitle
\section{Introduction}
Several types of urban radars have been researched and developed for civilian and military applications such as law enforcement, search and rescue, biomedical applications related to elderly monitoring and assisted living  \cite{amin2016through,amin2017radar,gurbuz2019radar,le2019radar,thayaparan2008micro,fioranelli2015classification}. 
The primary objectives of most indoor radars are human detection and localization. Moving humans are detected based on the Doppler modulations that are introduced to coherent radar transmit waveforms. The movements of the limbs of the human introduce micro-Doppler features which are captured through single dimension Doppler-time spectrograms using joint time-frequency transforms \cite{chen2003analysis,thayaparan2008micro,ram2008doppler,narayanan2010through,clemente2013developments}. When the continuous wave radar is augmented with array processing, we obtain Doppler enhanced images along azimuth \cite{ram2008through} or both azimuth and elevation \cite{lin2006frontal,ram2015high}. Alternatively, broadband pulse, linear frequency modulated or stepped frequency radars use fine downrange resolution to detect and track human motions along the range dimension \cite{ahmad2005synthetic,dehmollaian2008refocusing,ahmad2008three,yoon2008high}. The resulting radar signatures are either single dimension high range resolution profiles (range-time signatures) \cite{cammenga2015high} or range-enhanced higher order plots \cite{martone2009through,ahmad2013through}. Human activities are detected and interpreted on the basis of their micro-Doppler and micro-range features \cite{gennarelli2015multiple,he2015range,fogle2012micro, erol2019radar,gurbuz2019radar}.

Indoor radars could either be deployed in line-of-sight (LOS) environments (such as for fall monitoring of the elderly) or in non-line-of-sight (NLOS) environments (for security and surveillance purposes). The two most common NLOS deployments are the through-wall radar (TWR) \cite{amin2016through,amin2017radar,zhao2018through} and the around-the-corner (ACR) radar \cite{sume2009radar,sume2011radar,rabaste2015around,rabaste2017around,vishwakarma2020micro}. 
However, in both the cases, the quality of radar signatures are greatly impacted by complex propagation artifacts introduced by walls such as attenuation and multipath clutter \cite{ahmad2007autofocusing,leigsnering2014multipath,vishwakarma2018mitigation,vishwakarma2020mitigation}. We broadly categorise indoor clutter into two types- \emph{target independent}, and \emph{target dependent} static and dynamic clutter. 

Target independent static clutter arises from the reflections of radar signal off the lateral walls, ground, ceiling and other furniture in the rooms. While the target dependent static clutter is generated through reflections and refraction of signals from the target to side and back walls, reverberations within the front wall resulting in ghost targets and defocusing of targets. Several research efforts have been devoted to mitigate these artifacts on the radar signatures \cite{chen2016multipath,tan2014multipath}. The authors in \cite{martone2009through} and \cite{ahmad2013through} used back-projection and sparsity-based change detection algorithms, respectively, to track slow-moving humans in the range-crossrange space in the presence of target-dependent static clutter. These techniques relied on the availability of prior information of the wall geometry and characteristics in the indoor scenarios. Alternatively, authors in \cite{leigsnering2018sparsity,tang2017multipolarization} adopted the multipath exploitation strategy wherein the wall and target returns were projected on higher order subspaces based on their sparsity based representations. Then, the wall effects were removed from the target returns. While this is an effective strategy for removing target independent static clutter, it cannot be used for target dependent clutter since the target and clutter returns are no longer independent of each other. 
Dynamic clutter, on the other hand, arises due to the presence of other movers in the channel (target independent dynamic clutter) or due to the interactions between the dynamic target and the channel (target dependent dynamic clutter). The former can be separated on the basis of the micro-Doppler returns \cite{vishwakarma2017detection}. The latter is relatively difficult to remove as the target returns are not independent of complex wall propagation phenomenology.  

Besides clutter, indoor radar signatures are also affected by noise and interference. Indoor radars typically operate below X band frequencies in order to enable the radar signal penetration through the wall materials. These radars are federally mandated to transmit low powers to limit the possibility of electromagnetic interference with other wireless devices. On the other hand, these indoor radars may also encounter significant interference from neighboring wireless systems in the environment such as WiFi. Therefore, successful radar detection of targets in these circumstances, rely on effective clutter, noise and interference management strategies. In this paper, we propose to use denoising autoencoders (DAE) to recover high quality radar signatures similar to free space signatures even under low signal to clutter and noise ratios (SCNR).

\begin{figure}[htbp]
\centering \subfigure[]{
\includegraphics[width=2.5in, height=1.6in]{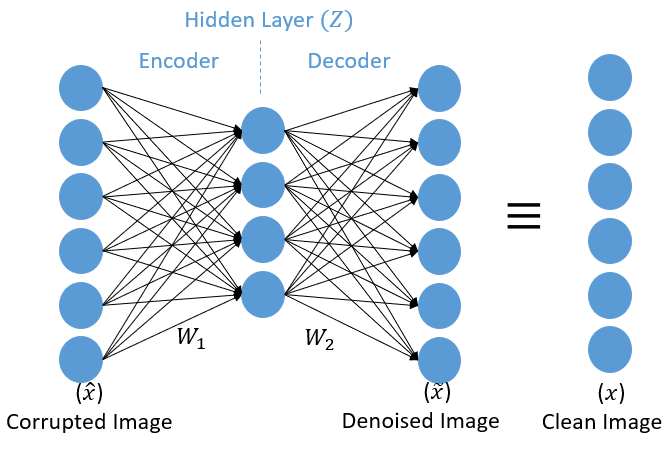}}
\\
\subfigure[]{
\includegraphics[width=3.3in, height=1.7in]{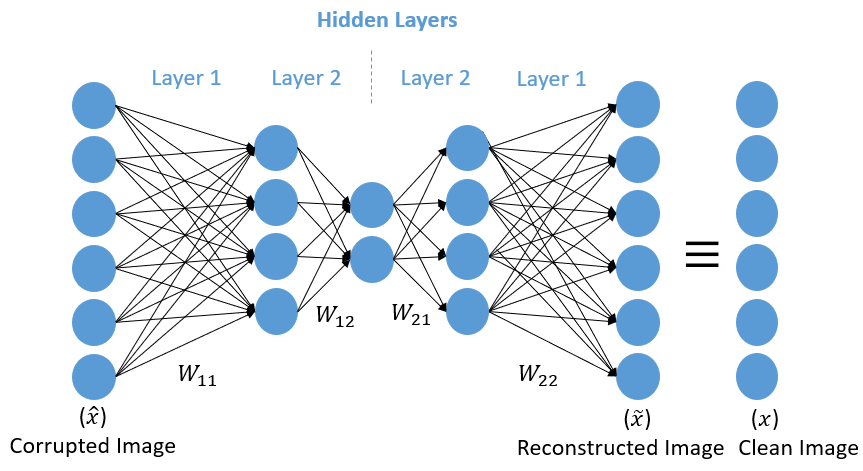}}
\caption{(a) Shallow denoising autoencoder and (b) Stacked denoising autoencoder}
\label{fig:denoising_autoencoder}
\end{figure}

DAE are neural networks with two stages - an encoder and a decoder as shown in Fig.\ref{fig:denoising_autoencoder}a \cite{DAE}.
During training, the encoder in the network is \emph{trained} to represent a noisy image ($\hat{x}$) with a hidden layer ($Z$). Then, the decoder is simultaneously \emph{trained} to recover a clean / denoised image ($\tilde{x}$) from the hidden layer that resembles a ground truth clean image ($x$). During test, the encoder is fed with noisy images, and clean images are gathered at the output of the decoder. The DAE has been widely applied in many different fields such as computer vision \cite{vincent2008extracting,hinton2006reducing}, anomaly detection and natural language processing \cite{sakurada2014anomaly,socher2011dynamic,vincent2008extracting,chen2012marginalized}. In \cite{vishwakarma2020mitigation}, we demonstrated the usefulness of DAE for removing dynamic clutter from through-wall frontal radar images. The algorithm demonstrated robust denoising and clutter mitigation performance for diverse wall and target conditions. However, the good performance of the network was predicated on two assumptions. First, the availability of a large volume of correctly labelled clean and noisy images during training. Second, the signal to noise ratio (SNR) of the radar system is high. However, in real world scenarios, both of these assumptions are often violated. For example, it may be nearly impossible to exactly replicate the same motion of the target in free space and then in the indoor channel conditions. Similarly, the channel may be plagued by high noise and interference, as discussed earlier. Therefore, in this paper, we propose to use a stacked DAE with sparsity constraints on the hidden layer representations \cite{ng2011sparse,DAE}, hitherto referred to as the StackedSDAE. In other words, instead of a a single hidden layer, we propose a cascade of hidden layers, as shown in Fig.\ref{fig:denoising_autoencoder}b, where the representations are generated from sparsity based constraints. The motivation of the approach is that such deeper representations enable the network to capture higher order abstractions in the clean and noisy images.
The incorporation of these additional hidden layers increases the computational time complexity during training but there is a significant reduction during \emph{test} due to the feature size reduction within the hidden layers. Preliminary studies with the StackedSDAE were presented in \cite{vishwakarma2019Clu}. In this paper, we offer more comprehensive experimental evaluation of the proposed algorithm.

The proposed StackedSDAE can be used for \emph{any} type of radar signature where corrupted and corresponding clean images are available. In this paper, we have tested the proposed algorithm on different types of narrowband and broadband radar images of a dynamic human. First, ACR signatures are simulated using a combination of electromagnetic modeling using a full wave solver and animation models of humans. Radar images are generated for different carrier frequencies and diverse electrical characteristics of the wall. The narrowband data are processed with short time Fourier transform (STFT) across the time domain to obtain Doppler-time spectrograms. The wideband data are similarly processed with Fourier transform across the carrier frequency bandwidth to obtain high range resolution profiles (HRRP). 
In the second scenario, we generate range enhanced frontal images of humans using wideband radar data augmented with two-dimensional array processing using an off-the-shelf sensor called the Walabot  \cite{walabot_2017}. All of these images are corrupted by clutter and noise and the StackedSDAE is evaluated on them. 
Our results, across all three types of images, show that the StackedSDAE can considerably mitigate the clutter and distortions introduced by the walls even with high labelling mismatch error and under low SNR. 

To summarize, the main contributions of our paper are the following.
First, we propose StackedSDAE for mitigating clutter and distortions introduced by different types of wall phenomenology. The framework of the proposed algorithm can be used to denoise any type of distorted radar signature.
Second, we have performed experimental validation on three different types of radar signatures - the Doppler time spectrogram, the HRRP and the range enhanced frontal images. For this purpose, we have generated a large database of these radar signatures in diverse channel conditions and radar frequencies. The complete database and algorithms are shared with the research community on the following URL \url{https://rb.gy/mmhzf6}. Third, we have shown that the proposed algorithm is particularly effective  under low SNR conditions and when there are large errors in the labeling of the training data.

Our paper is organized in the following manner. In section \ref{Sec:Theory}, we discuss the theory of StackedSDAE for clutter mitigation of radar images. In the subsequent section, we describe the experimental data generation through simulations followed by the simulation results in section 4. In section 5, we describe the measurement data collection followed by the denoising results from the autoencoder. We conclude with the analysis of the results in the final section. 

\emph{Notation:} In this paper, we represent scalars and column vectors by lower case letters and matrices by upper case letters. 
\section{Theory}
\label{Sec:Theory}
When radars are operated in highly cluttered environments, significant distortions arise in the images. The DAE is a neural network that can be trained to remove the clutter artefacts from these images. The autoencoder requires both clean and noisy images while training. In the case of radar, the clean images correspond to radar images of the target in free space or some environment free of the clutter artefacts. These are denoted by $X\in \Re ^{P\times Q}$ where $P$ is the pixel size of each of the $Q$ images. In other words each image, $x$, of $P$ pixels is vectorized and then $Q$ such images are stacked column-wise to generate $X$.  The corrupt data, $\hat{X}\in \Re ^{P\times Q}$ are corresponding images of a similar target undertaking a similar motion, gathered in cluttered environments. In this section, we describe the DAE and two variants, the sparsity based DAE - termed as SparseDAE - and the sparsity based stacked DAE - termed as StackedSDAE. Our hypothesis is that these variants will outperform the conventional DAE in clutter mitigation.  
\subsection{Denoising Autoencoder (DAE)}
We begin with a description of the conventional / standard single layer DAE framework, shown in Fig.\ref{fig:denoising_autoencoder}(a). Here, the algorithm learns a weight matrix, $W_{1}\in \Re ^{l\times Q}$, in order to represent $\hat{X}^{tr}$ with a compressed $Z\in \Re ^{l\times Q}$ through 
\begin{equation}
\label{eqn:encode}
    Z=\phi(W_1\hat{X}^{tr}).
\end{equation}
The number of nodes, $l$, in the weight matrix are fewer than the original pixel size of the image. 
In the decoder, $Z$ is mapped back to the reconstructed clean image, $\tilde{X}^{tr}$ through
\begin{equation}
\label{eq:recovery}
\tilde{X}^{tr}=W_2 \phi (W_1 \hat{X}^{tr}),   
\end{equation}
where $W_2$ is the weighting function. In both the encoder and decoder, the same activation function $\phi$, is used which could be either linear or a non-linear. Some of the popular non-linear activation functions in literature are hyperbolic tangent and sigmoid \cite{mehta2016stacked,MAJUMDAR2018271}.  
The objective of the algorithm is to learn $W_1$ and $W_2$ from the training data such that the normalized mean square error between $X^{tr}$ and $\tilde{X}^{tr}$ is minimized as shown in
\begin{equation}
\label{eqn:objective_func} 
    \min_{W_1,W_2} \left \|  X^{tr}-W_2\phi(W_1\hat{X}^{tr})\right \|_{2}^{2}.
\end{equation}
Equation (\ref{eqn:objective_func}) is a complex optimization problem which is NP hard to solve. Therefore, we introduce a proxy variable, $Z$, as shown in 
\begin{equation}
  \label{eqn:ObjFunc1} 
    \min_{W_1,W_2} \left \|  X^{tr}-W_{2}Z\right \|_{2}^{2} \hspace{0.1cm}
    s.t.\hspace{0.1cm} Z=\phi(W_1\hat{X}^{tr}).
\end{equation}
Then we relax the equality constraint in the formulation using an augmented Lagrangian, $\lambda$, in 
\begin{equation}
\label{eqn:ObjFunc2} 
      \min_{W_1,W_2,Z} \left \|  X^{tr}-W_2Z\right \|_{2}^{2}
    + \lambda\left \| Z-\phi(W_1\hat{X}^{tr})\right \|_{2}^{2}. 
\end{equation}
The regularization parameter, $\lambda$, in the above expression trades off between the error in the encoder (second term) and the decoder (first term) stages.  
The above formulation has a close form solution using an alternating direction method of multipliers (ADMM) \cite{boyd2011distributed}. In the following section, we will describe the implementation of the ADMM in greater detail. 

Once trained, the DAE is ready for test. During test, a denoised radar image $\tilde{x}^{test}$ is recovered from the DAE with the test noisy image $\hat{x}^{test}$ as input using \eqref{eq:recovery}. The denoised image should now resemble the ground reference clean image $x^{test}$.
\subsection{Sparse Denoising Autoencoder (SparseDAE)}
A modified autoencoder framework can be derived by imposing additional sparsity constraints on the hidden layer representations ($Z$) while learning the weighting matrices in the encoder and decoder. The objective function in \eqref{eqn:ObjFunc2} is modified to  
\begin{equation}
\label{eqn:ObjFunc3} 
      \min_{W_1,W_2,Z} \left \|  X^{tr}-W_2Z\right \|_{2}^{2}
    + \lambda\left \| Z-\phi(W_1\hat{X}^{tr})\right \|_{2}^{2}+\mu\left | Z \right |_1, 
\end{equation}
where an $l_1$ norm has been imposed on $Z$ through a second regularization parameter $\mu$. The objective function is solved through ADMM by separately solving for $W_1$, $W_2$ and $Z$ through iterations.
First, $W_1$ is obtained from the closed form least squares solution for 
\begin{equation}
\label{eq:P1}
 \min_{W_1}\left \| \phi^{-1}Z-W_1\hat{X}^{tr}\right \|_{2}^{2}.   
\end{equation}
\label{eq:P2}
Then $W_2$ is similarly solved using least squares in
\begin{equation}
\min_{W_2}\left \|  X^{tr}-W_2Z\right \|_{2}^{2}.
\end{equation}
Then both $W_1$ and $W_2$ are used to solve for $Z$ in
\begin{equation}
\label{eq:P3}
\begin{split}
\min_{Z} \left \|  X^{tr}-W_2Z\right \|_{2}^{2}+ \lambda\left \| Z-\phi(W_1\hat{X}^{tr})\right \|_{2}^{2}+\mu\left | Z \right |_1 \\
= \min_{Z} \begin{Vmatrix}
\begin{pmatrix}
X^{tr}\\ 
\sqrt{\lambda}\phi(W_1\hat{X}^{tr})
\end{pmatrix}-\begin{pmatrix}
W_2\\ 
\sqrt{\lambda}I
\end{pmatrix}{Z}
\end{Vmatrix}^{2}_{2}+\mu\left | Z \right |_1, 
\end{split}
\end{equation}
using iterative soft thresholding algorithm (ISTA) \cite{bishop2006pattern}. 
We update the network weight $W_1$, $W_2$ and proxy variable $Z$, iteratively until the algorithm converges. 

Once the network is trained, the weight matrices $W_1$ and $W_2$ are used obtain a denoised form $\tilde{x}^{test}$ of the corrupted test data $\hat{x}^{test}$ using 
\begin{equation}
\label{eqn:test_stage1}
 \tilde{x}^{test}=W_2\phi(W_1\hat{x}^{test}).   
\end{equation}
The hypothesis is that when the autoencoder is properly trained the error after denoising (AD) between $\tilde{x}^{test}$ and $x^{test}$, is lower than the error \emph{before denoising} (BD) between $\hat{x}^{test}$ and $x^{test}$ where $x^{test}$ is the corresponding ground truth clean image.
\subsection{Stacked Sparse Denoising Autoencoder (StackedSDAE)}
In this framework, the single hidden layer representation within the autoencoder is converted to multiple stacked layers as shown in Fig.\ref{fig:denoising_autoencoder}b. There is a vast body of research that have demonstrated that additional deeper layers in a neural network enable capturing of higher order abstractions in the data resulting in significant improvement in the performance of the algorithms. This is because the successive layers result in reuse of key features within the images as well as extraction of higher order features. The number of the layers and the number of nodes within each layer are typically heuristically chosen.
In our work, we implement the StackedSDAE using three hidden layers. Therefore, instead of learning just two weighting matrices as was the case of the shallow DAE, our objective, here is to learn $W_{11},W_{12},W_{21}$ and $W_{22}$.  Each succeeding deeper layer is characterized by fewer number of nodes. As a result, the computational time complexity increases during the training phase since there are greater number of training matrices to learn. However, the complexity during the test phase reduces because of the reduced feature dimensions of the stacked layers.  

Again, we divide our denoising problem into the training and the test stages. During training, the denoising problem can be formulated as  
\begin{gather}
\label{eqn:objective_func4} 
    \min_{W_{11},W_{12},W_{21},W_{22}} \nonumber\\
    \left \|  X^{tr}-W_{22}\phi \left( W_{21}\phi \left( W_{12}\phi \left( W_{11}\hat{X}^{tr} \right) \right) \right)\right \|_{2}^{2}.
\end{gather}
\\
\noindent Again the problem is NP hard since it is non-convex. Similar to SparseDAE, we use the variable separation technique by introducing proxy variables $Z_2$, $Z_1$ and $Z_0$ such that
\begin{align}
   Z_2=\phi(W_{21}Z_1) \\
   Z_1=\phi(W_{12}Z_0) \\
   Z_0=\phi(W_{11}\hat{X}^{tr}) 
\end{align}
Upon relaxing these constraints with augmented Lagrangian, the objective function now becomes
\begin{align}
\label{eqn:ObjFunc2d} 
\begin{split} 
   & \min_{W_{11},W_{12},W_{21},W_{22},Z_{2},Z_{1},Z_{0}} \left \|  X^{tr}-W_{22}Z_{2}\right \|_{2}^{2}\\
   &\hspace{9ex} +\mu_2 \left\|\Phi ^{-1}(Z_2)-W_{21}Z_1\right \|_{2}^{2}\\
    & \hspace{9ex}+\mu_1 \left\|\Phi ^{-1}(Z_1)-W_{12}Z_0\right \|_{2}^{2}\\
     &\hspace{9ex} +\mu_0 \left\|\Phi ^{-1}(Z_0)-W_{11}\hat{X}^{tr}\right \|_{2}^{2}
\end{split}
\end{align}
Again, we use the ADMM technique for solving the above formulation. We separately solve for $W_{11}, W_{12}, W_{21}$ and $W_{22}$, using closed form expressions for least squares,  as shown below
\begin{eqnarray}
\label{eqn:W11}
 \min_{W_{11}}\left \| \phi^{-1}(Z_0)-W_{11}\hat{X}^{tr}\right \|_{2}^{2}, \\
 \label{eqn:W12}
 \min_{W_{12}}\left \| \phi^{-1}(Z_1)-W_{12}Z_0\right \|_{2}^{2},  \\
 \label{eqn:W21}
 \min_{W_{21}}\left \| \phi^{-1}(Z_2)-W_{21}Z_1\right \|_{2}^{2}, \text{ and}\\
 \label{eqn:W22}
\min_{W_{22}}\left \| X^{tr}-W_{22}Z_2\right \|_{2}^{2}.
  \end{eqnarray}
Then using the ISTA algorithm and the weighting matrices, we solve for $Z_0$, $Z_1$ and $Z_2$ based on the following objective functions:
\begin{eqnarray}
\label{eq:Z0}
\min_{Z_0} \left \|  \phi^{-1}Z_1-W_{12}Z_0\right \|_{2}^{2}+ \mu_0\left \| Z_0-\phi(W_{11}\hat{X}^{tr})\right \|_{2}^{2} \nonumber\\
+\lambda_0\left | Z \right |_1, \\
\label{eq:Z1}
 \min_{Z_1} \mu_2\left \|  \phi^{-1}Z_2-W_{21}Z_1\right \|_{2}^{2}+ \mu_1\left \| Z_1-\phi(W_{12}Z_0)\right \|_{2}^{2} \nonumber\\
 +\lambda_1\left | Z \right |_1, \text{ and} \\
 \label{eq:Z2}
\min_{Z_2} \left \|  X^{tr}-W_{22}Z_2\right \|_{2}^{2}+ \mu_2\left \| Z_2-\phi(W_{21}Z_1)\right  \|_{2}^{2}\nonumber\\
+\lambda_2\left | Z \right |_1. 
\end{eqnarray}
Equations \eqref{eqn:W11} to \eqref{eq:Z2} are iterated till the algorithm converges. 

Once the network weights are trained, we use them to reconstruct $\tilde{x}^{test}$ from the corrupted $\hat{x}^{test}$ by
\begin{equation}
\label{eqn:test_stage2}
 \tilde{x}^{test}=W_{22}\phi(W_{21}\phi(W_{12}\phi(W_{11}\hat{x}^{test}))).
\end{equation}
Note that the StackedSDAE algorithm is significantly faster in generating denoised images at test time as it involves only a simple product operation with reduced feature dimensions than DAE and SparseDAE. This makes the algorithm suitable for real-time applications where training is usually done apriori.
\subsection{Evaluation Metrics}
The objective of the DAE and its variants are to reconstruct radar images that resemble those that would be obtained if the target were to move in free space conditions. Therefore, one obvious metric is the normalized mean square error between the reconstructed image ($\tilde{x}^{test}$) which is obtained after denoising and the ground truth free space image ($x^{test}$). However, in the image processing community, other metrics are preferred to NMSE since NMSE does not compare the salient features between different images. In this paper, we use the structural similarity index (SSIM) which is a popular metric that assesses luminance, contrast and structural differences between two images \cite{wang2004image}. Its value ranges from 0 to 1 where 1 is obtained when the images are identical.
We calculate SSIM between the cluttered image $\hat{x}^{test}$ and $x^{test}$ before denoising (BD). Then the  SSIM is calculated between  $\tilde{x}^{test}$ and $x^{test}$ after denoising (AD). The hypothesis, here, is that the SSIM will approach unity after denoising.
\section{Simulation Methodology}
We test the denoising algorithms on indoor radar signatures obtained in NLOS scenarios where there is both noise and clutter. We specifically consider the ACR scenario for simulations study as described below.
\subsection{Simulation Models}
\begin{figure}[htbp]
\centering
\includegraphics[scale=0.35]{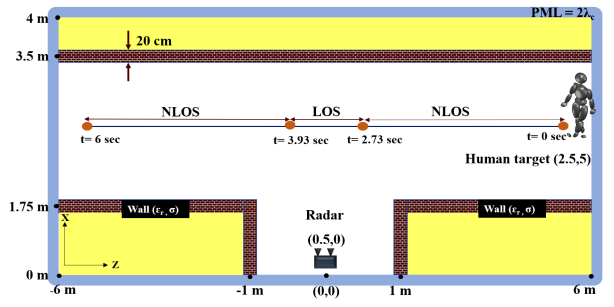}
\caption{Simulation model of 6 seconds long human walking motion in around-the-corner radar (ACR) scenario in the presence of 20cm thick walls with dielectric constant of $\epsilon_r$ and conductivity of $\sigma$. Wave propagation from the radar source at (0.5,0)m modeled using two-dimensional full wave electromagnetic solver.}
\label{fig:ACR_Setup}
\end{figure}

The ACR simulation set up is shown in Fig.\ref{fig:ACR_Setup}. 
The electromagnetic wave propagation from the radar is modeled using two-dimensional finite difference time domain (FDTD) simulations in the $XZ$ space. The simulation space consists of two corridors of 2m width arranged in a T-shape as shown in the figure. The walls are assumed to be 20cm thick. The simulation space is discretized to form uniform grid cells that are a tenth of a wavelength ($\lambda_c$) of the carrier frequency ($f_c$). All the regions outside of the walls are assumed to be of free space. Stochasticity is introduced in the electrical characteristics of the walls. Each grid cell within the wall has a dielectric constant that is drawn from a normal distribution, $\mathcal{N}(\epsilon_r,\epsilon_r^{std})$, of mean $\epsilon_r$ and a standard deviation of $\epsilon_r^{std}$. Similarly, the conductivity of each grid cell is drawn from the normal distribution of $\mathcal{N}(\sigma,\sigma^{std})$. Therefore, the walls are not truly homogeneous since each grid cell has slightly different electrical characteristics to model real world conditions. The simulation space is bounded by a perfectly matched layer that is $2\lambda_c$ thick.  
The source excitation which models the monostatic radar is located at $(0.5,0)m$. Two types of source excitation are considered. The first is a narrowband source modelled as a sinusoidal source of $f=f_c$ frequency. The second is a broadband source excitation ($f=f_c \pm \frac{\beta}{2}$) which is modelled as a Gaussian signal modulated by the sinusoidal carrier signal at $f_c$. The width of the Gaussian signal determines the bandwidth ($\beta$) of the source excitation. The time-domain simulations are allowed to run long enough to ensure that steady state conditions are reached and the  mean and standard deviation of the time-domain electric field at every point in the simulation space are saved \cite{smith2012stochastic}. Based on the normal distribution, multiple realizations ($\eta = 1: M$) of the time-domain electric field at each two-dimensional grid position, $\vec{\rho},$ are generated. Each of these $\eta$ electric field vectors are then transformed using Fourier transform to the frequency domain and complex responses, $H(\vec{\rho},f,\eta)$, at $f=f_c \pm \frac{\beta}{2}$, are saved.

Next, we consider the human moving along the tangential trajectory before the radar, as shown in the figure, over a duration of T seconds. The human is a three dimensional figure with the height along $Y$ axis. The skeleton framework of the human and the animation motion of the body parts are described using motion capture data from Sony \cite{ram2008simulation}. Then the electromagnetic radar scattering off the human are modeled using the techniques described in \cite{ram2010simulation}. We briefly describe the technique here. The human is considered to be a collection of $B$ discrete point scatterers corresponding to different body parts each of $a_b$ reflectivity. Each of the body parts is modelled as an ellipsoid whose radar cross-section ($a_b^2$) is obtained with analytical expressions. The time-domain radar returns from the human, corresponding to each $\eta$ stochastic FDTD realization at frequency $f$,  are obtained by
\begin{equation}
\label{eq:hybridem}
\begin{split}
s_{rx}(t,f,\eta) = \sum_{b=1}^B A a_b H(\vec{\rho}_b(t),f,\eta)^2e^{-j4\pi \frac{f}{c} (r_b (t)- \rho_b(t)) }, \\
t=0:T, f=f_c\pm \beta, \eta = 1:M.
\end{split}
\end{equation}
Here $r_b(t)$ and $\rho_b(t)$ are the time-varying three and two-dimensional Euclidean distances of the $b^{th}$ point scatterer from the radar respectively. The two way propagation physics from the radar to the point scatterer is captured by the square of the wall response $H$. Since the FDTD is a two-dimensional simulation with an infinite line source excitation, the exponential phase term in \eqref{eq:hybridem} corrects the circular phase front from the two-dimensional FDTD propagation physics to the spherical phase front in the three-dimensional scenario. $A$ in the above expression calibrates the amplitude of the FDTD source excitation to desired radar equivalent isotropic radiated power.  

The time-domain radar data, $s_{rx}(t,f,\eta)$, could be narrowband or wideband. In the case of narrow-band data (where $\beta =0$), the short-time Fourier transform is applied on the data to obtain M Doppler spectrograms, $x_{DT}$, as shown in
\begin{equation}
    x_{DT}(\tau,f_D,\eta) = \int_t s_{rx}(t,\eta)w(\tau-t)e^{-j2\pi f_D t}dt, \eta = 1:M,
\end{equation}
where $w(t)$ is the short time window function. 
In the case of the broadband data, Fourier transform is applied across the bandwidth of the data at every time instant $t$ to obtain M high range resolution profile (HRRP), $x_{HRRP}$, as shown in
\begin{equation}
    x_{HRRP}(t,r,\eta) = \int_{f} s_{rx}(t,f,\eta)e^{+j 2\pi f \frac{2r}{c} } df, \eta = 1:M.
\end{equation}

\subsection{Simulated Radar Signatures}
\emph{Narrowband Doppler-time Signatures:}
The above process is carried out independently for three radar carrier frequencies - $f_c:$ 2.4GHz, 5GHz and 10GHz - and the radar bandwidth is set at 0Hz. For each of the above cases, we consider three different types of walls. A wall with low mean conductivity $\sigma = 0.05S/m$, medium conductivity ($\sigma = 100S/m$) and high conductivity ($\sigma = 1e5S/m$). The mean dielectric constant is fixed at $\epsilon_r = 4$. The standard deviation for both dielectric constant and conductivity are fixed at 30\%. Twenty stochastic realizations of the time-domain electric field are generated for each of the three cases using the stochastic FDTD solver. These realizations are combined with the human walking motion to generate ACR Doppler-time signatures. A short time window of 0.1s is used to generate the spectrograms. The total duration of the human motion is 6s. This interval is separated into 8 consecutive intervals of 0.75s duration. The sampling frequency of the time domain data is 500Hz resulting in Doppler frequency axes spanning from $f_D = -250Hz : +250Hz$ in all the images. Then complex Gaussian additive noise, $\mathcal{N}(0,N_p)$, of $N_p$ noise power are added to each pixel of the images to realize 200 images. Therefore we have a total of 1600 images for each of the three wall cases.

\begin{figure*}[ht]
\centering
\includegraphics[scale=0.7]{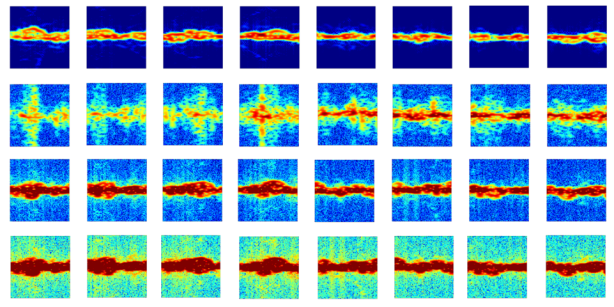}
\caption{Doppler-time spectrograms of human walking in ACR scenario with monostatic narrowband radar operating at \textbf{2.4GHz}. The y-axis across all figures shows Doppler frequencies spanning from -250Hz to +250Hz. The x-axis across each figure is of 0.75s duration with a total time duration of 6s across the 8 columns. The dynamic range of each figure is from -20 to -70dB. The first, second, third and fourth rows show the radar signature of human walking in free space, low conductive, medium conductive and high conductive wall conditions, respectively. {SNR for all figures along bottom three rows is fixed at $-20dB$. }}
\label{fig:STFT_2pt4GHz}
\end{figure*}

We show the wall propagation effects on the images at 2.4GHz in Fig.\ref{fig:STFT_2pt4GHz}.
The top row on the figure shows the results generated in free space conditions in the absence of clutter and noise. These are generated when the human walks in the presence of radar but without any walls. The resulting micro-Doppler spectrograms across 8 time intervals from 0 to 6 seconds are shown. We observe that the Dopplers are low since the human motion is tangential with respect to the radar. As the human approaches the radar, from 0s to approximately 3s, the Dopplers are positive. Then, when the human moves away, the Dopplers become negative. We observe weak micro-Doppler returns from the other body parts. The second row shows the returns when the human is walking in the presence of low conductive walls. In this scenario, the walls allow the signal to penetrate with some attenuation. Hence, the strength of the signals are weaker. Due to the multipath introduced by the ringing of the radar signal within the wall, we observe a lot of multipath. From this figure, it becomes difficult to know whether there are one or more targets moving and whether they are coming towards or away from the radar. The third and fourth row show the micro-Dopplers from a medium and a high conductive wall. Due to the lossy nature of the walls, the through-wall propagation is blocked. Instead, the dominant phenomenon here is the reflections off the lateral walls which give rise to high strengths in the radar scattered signal. There is lesser micro-Doppler spread, in these cases. However, we do observe some negative Dopplers due to multipath, even when the target approaching the radar (the first few columns).     

\begin{figure*}[ht]
\centering
\includegraphics[scale=0.7]{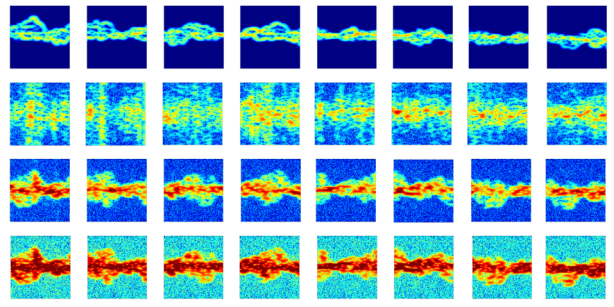}
\caption{Doppler-time spectrograms of human walking in ACR scenario with monostatic narrowband radar operating at \textbf{5GHz}. The y-axis across all figures shows Doppler frequencies spanning from -250Hz to +250Hz. The x-axis across each figure is of 0.75s duration with a total time duration of 6s across the 8 columns. The dynamic range of each figure is from -20 to -70dB. The first, second, third and fourth rows show the radar signature of human walking in free space, low conductive, medium conductive and high conductive wall conditions, respectively. {SNR for all figures along bottom three rows is fixed at $-20dB$.}}
\label{fig:STFT_5GHz}
\end{figure*}

Next, we present the results for the 5GHz carrier in Fig.\ref{fig:STFT_5GHz}.
The figures show a slightly lower strength compared to the results from 2.4GHz due to the antenna gain offset between the two frequencies.
The higher carrier frequency results in finer Doppler resolution. As a result, we are able to discern distinct micro-Doppler tracks from the different body parts in the free space scenario, shown in the top row. The low conductive walls give rise to significant through-wall propagation resulting in micro-Doppler spread and radar signal attenuation. This results in the low clarity spectrograms in the second row. Again, the third and fourth row show that the through-wall propagation has been blocked. However, multipath reflections off the lateral walls give rise to negative Dopplers even when the target Doppler is positive with respect to radar. 

\begin{figure*}[ht]
\centering
\includegraphics[scale=0.7]{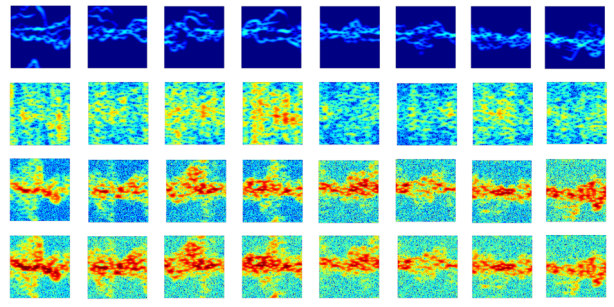}
\caption{Doppler-time spectrograms of human walking in ACR scenario with monostatic narrowband radar operating at \textbf{10GHz}. Again, the signals are weaker here due to the antenna gain offset. The y-axis across all figures shows Doppler frequencies spanning from -250Hz to +250Hz. The x-axis across each figure is of 0.75s duration with a total time duration of 6s across the 8 columns. The dynamic range of each figure is from -20 to -70dB. The first, second, third and fourth rows show the radar signature of human walking in free space, low conductive, medium conductive and high conductive wall conditions, respectively.{SNR for all figures along bottom three rows is fixed at $-20dB$.}}
\label{fig:STFT_10GHz}
\end{figure*}

In the Doppler-time spectrograms corresponding to 10GHz, in Fig.\ref{fig:STFT_10GHz}, we observe well resolved micro-Doppler tracks from the different body parts in the free space scenario. 
Due to the low sampling frequency we also observe some aliasing at negative frequencies even for the free space scenario (first few figures along top row). The wall distortions are again considerable due to both through-wall propagation effects (for low conductive walls) and due to multipath off lateral walls (for the high conductive walls). 

\begin{figure*}[ht]
\centering
\includegraphics[scale=0.7]{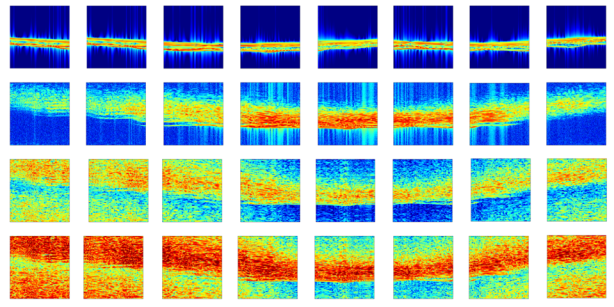}
\caption{High-range resolution profile of human walking in ACR scenario with monostatic narrowband radar operating at \textbf{2.4GHz}. The y-axis across all figures {shows range spanning from 0m to 10m.} The x-axis across each figure is of 0.75s duration with a total time duration of 6s across the 8 columns. The dynamic range of each figure is from -30 to -70dB. The first, second, third and fourth rows show the radar signature of human walking in free space, low conductive, medium conductive and high conductive wall conditions, respectively.{SNR for all figures along bottom three rows is fixed at $-20dB$.}}
\label{fig:HRRP_2pt4GHz}
\end{figure*}

\emph{Broadband Doppler-time Signatures:} We simulated broadband radar data of 2GHz bandwidth about the three carrier frequencies using the FDTD solver. This results in a range resolution of 0.075m and the maximum unambiguous range, based on the frequency step size, is 10m. Complex Gaussian noise was added to the pixels of the resulting HRRP images and a total of 1600 images were generated for each wall type. 
We first discuss the HRRP obtained at 2.4GHz in Fig.\ref{fig:HRRP_2pt4GHz}. 
Again, the total duration of the target motion is 6s and divided into 8 intervals of 0.75s each. The top row shows the target motion in free space conditions (in the absence of walls). We observe the range of the target changing only slightly as the human is moving tangentially across the radar's field-of-view. We are able to observe fine micro-range tracks arising from the motion of the limbs. However, the images significantly deteriorate in the presence of the walls due to multipath. In the case of the low conductive wall (second row), the ringing of the signal through the wall gives rise to multipath but also attenuates the radar signal. For high, conductive walls, the multipath arises due to reflections off the lateral walls which cause the radar received signal strength to increase. We also observe considerable aliasing in these scenarios. 

\begin{figure*}[ht]
\centering
\includegraphics[scale=0.7]{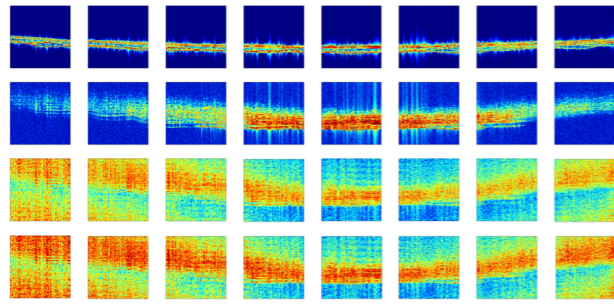}
\caption{High-range resolution profile of human walking in ACR scenario with monostatic narrowband radar operating at \textbf{5GHz}. The y-axis across all figures shows {range spanning from 0m to 10m.} The x-axis across each figure is of 0.75s duration with a total time duration of 6s across the 8 columns. The dynamic range of each figure is from -40 to -80dB. The first, second, third and fourth rows show the radar signature of human walking in free space, low conductive, medium conductive and high conductive wall conditions, respectively.{SNR for all figures along bottom three rows is fixed at $-20dB$.}}
\label{fig:HRRP_5GHz}
\end{figure*}

\begin{figure*}[ht]
\centering
\includegraphics[scale=0.7]{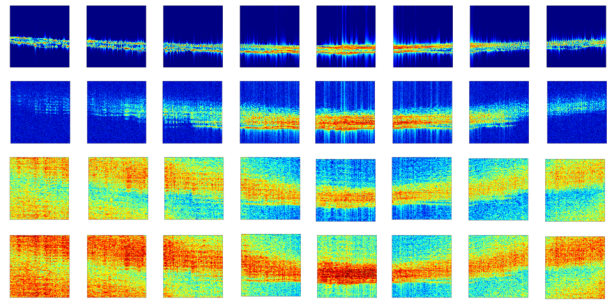}
\caption{High-range resolution profile of human walking in ACR scenario with monostatic narrowband radar operating at {10GHz}. The y-axis across all figures shows {range spanning from 0m to 10m.} The x-axis across each figure is of 0.75s duration with a total time duration of 6s across the 8 columns. The dynamic range of each figure is from -50 to -90dB. The first, second, third and fourth rows show the radar signature of human walking in free space, low conductive, medium conductive and high conductive wall conditions, respectively.{SNR for all figures along bottom three rows is fixed at $-20dB$.}}
\label{fig:HRRP_10GHz}
\end{figure*}

Similar phenomena are observed in the HRRP for 5GHz and 10GHz, which are shown in Fig.\ref{fig:HRRP_5GHz} and Fig.\ref{fig:HRRP_10GHz}. 
Again, the top row in both these figures show the HRRP when the human is walking in free space conditions where there are no walls and no noise. 
The second, third and fourth rows show the results when the walls are of low, medium and high conductivity respectively. The HRRP show a lot of similarity across the three carrier frequencies. This is mainly because the HRRP features are a function of the range resolution and the bandwidth of the radar which are identical across the three cases. The results from 10GHz show the greatest distortions and clutter. 
\section{Simulation Results and Analysis}
\begin{figure}[htbp]
\centering
\includegraphics[scale=0.35]{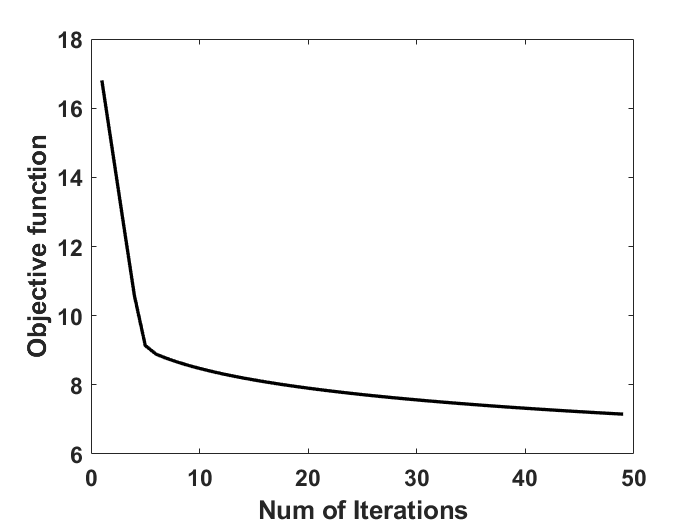}
\caption{{Convergence curve of the objective function for the denoising algorithm for a fixed SNR of 0dB.}}
\label{fig:Obj_func}
\end{figure}

In this section, we compare the similarity of the denoised / reconstructed images obtained from DAE, SparseDAE and StackedSDAE with respect to the clean ground truth images through the SSIM metric. The performances are evaluated for both types of radar signatures - the Doppler-time spectrograms and the HRRP - that were discussed above. We consider two parameters for comparison - the labelling mismatch error and the SNR, which is the ratio of the minimum signal receivable by the radar to the mean noise floor. In real world conditions, it may be impossible to exactly replicate a target motion in free space and ACR conditions during training. Therefore, the training data may have significant mismatch between the clean $X^{tr}$ and the noisy and cluttered $\hat{X}^{tr}$. We model this training error by shuffling the row entries of each column of $X^{tr}$ such that they no longer exactly correspond to the entries in $\hat{X}^{tr}$. We used 70\% of total images as the training data set, and the remaining 30\% as the test data set. We present the convergence of the objective function (shown in \ref{eqn:ObjFunc2}) with the number of iterations in Fig. \ref{fig:Obj_func}. We then change the labelling mismatch error percentage from 0 to 60\% by changing the degree of shuffling. Next, we change the SNR of the images from $-15dB$ to $+20dB$ by changing the Gaussian noise power $N_p$ that is added to each pixel of the image.  
\subsection{Time-Frequency Spectrograms}
\begin{figure*}[ht]
\centering 
\subfigure[]{
\includegraphics[scale=0.30]{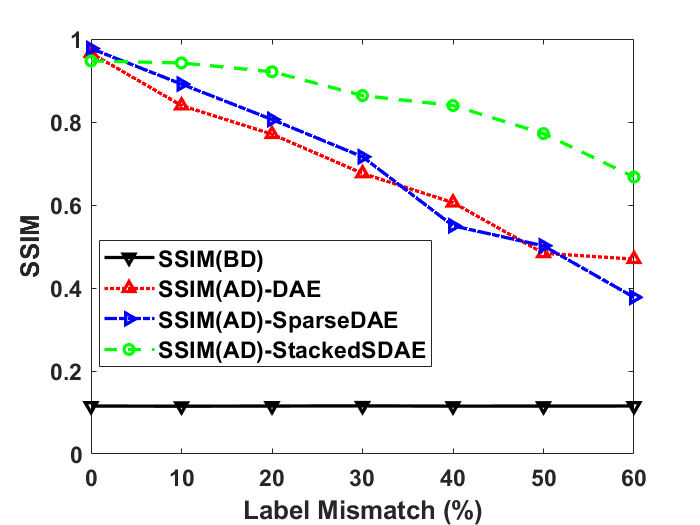}}
\subfigure[]{
\includegraphics[scale=0.30]{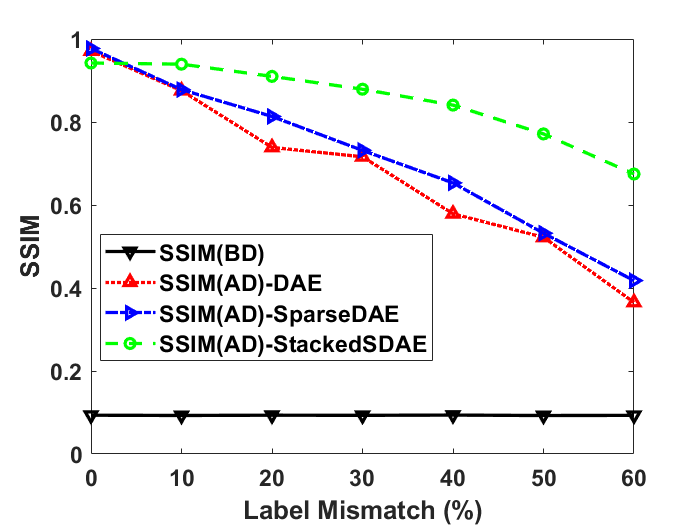}}
\subfigure[]{
\includegraphics[scale=0.30]{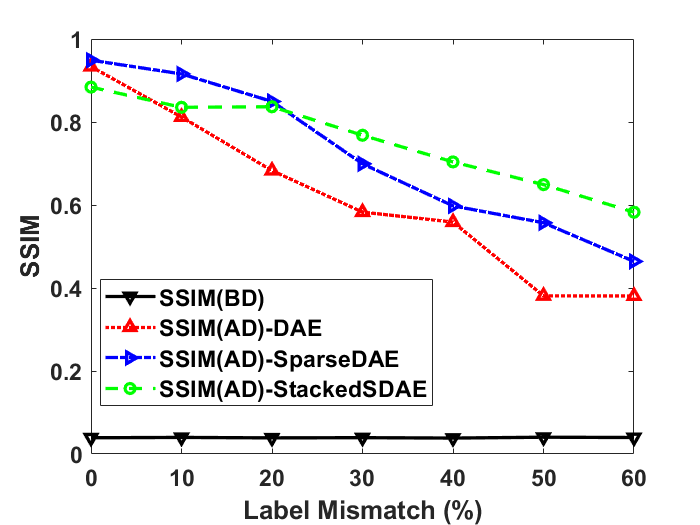}}
\caption{ SSIM variation for spectrograms with respect to label mismatch percentage at (a) 2.4GHz, (b) 5GHz, and (c) 10GHz respectively for fixed SNR of $-10dB$. }
\label{fig:STFT_labellingmismatch}
\end{figure*}

Fig.\ref{fig:STFT_labellingmismatch} shows the results of the three algorithms for different labelling mismatch errors across the three carrier frequencies at SNR of $-10dB$. 
In each case, the SSIM after denoising (AD) for all three algorithms - DAE, SparseDAE and StackedSDAE - are significantly improved when compared to the noisy and cluttered images before denoising (BD). As the label mismatch error increases, the SSIM for all three algorithms fall. However, the fall is much lower for the StackedSDAE. Even with 50\% labelling mismatch error, the SSIM of the reconstructed images are at 0.8 at 2.4GHz (Fig.\ref{fig:STFT_labellingmismatch}a) and 5GHz (Fig.\ref{fig:STFT_labellingmismatch}b). However, the 10GHz scenario is a far more challenging case, where we observed a lot of multipath from lateral walls and weaker signals due to antenna gain offset. However, even here, the SSIM of the StackedSDAE is better than the SparseDAE and DAE at high labelling errors.

\begin{figure*}[ht]
\centering 
\subfigure[]{
\includegraphics[scale=0.30]{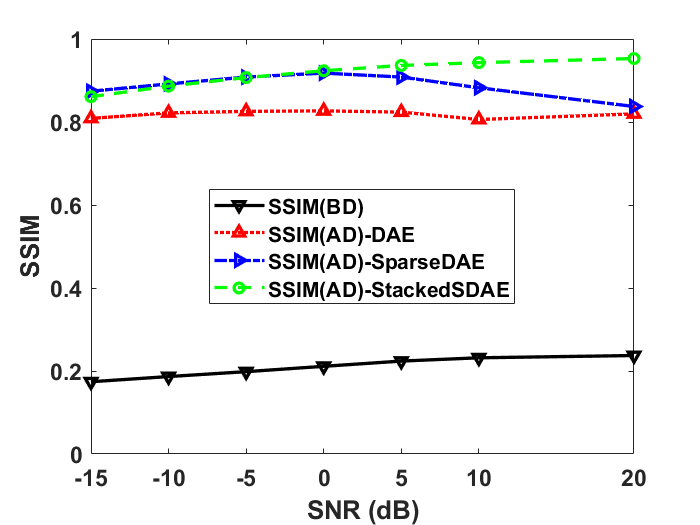}}
\subfigure[]{
\includegraphics[scale=0.30]{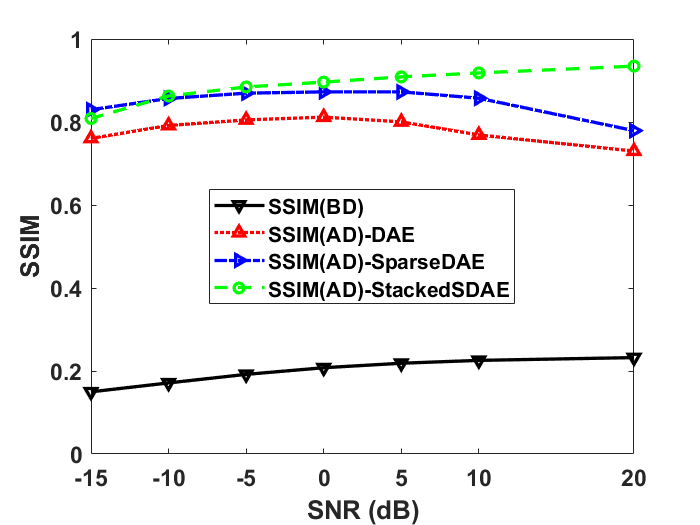}}
\subfigure[]{
\includegraphics[scale=0.30]{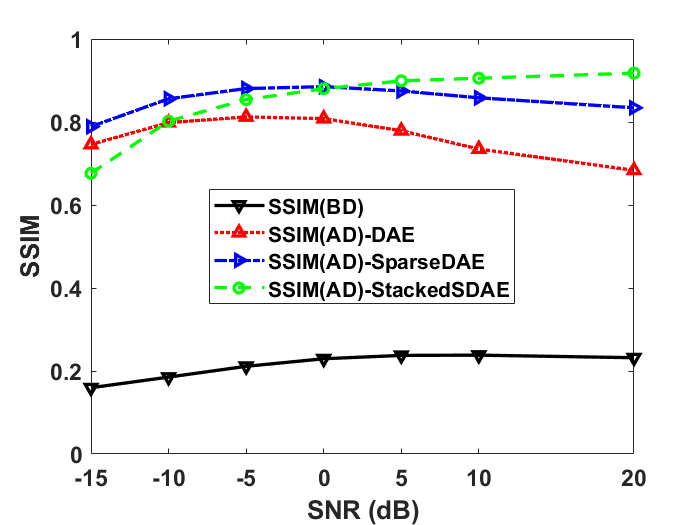}}
\caption{ SSIM variation for spectrograms with respect to SNR at (a) 2.4GHz, (b) 5GHz, and (c) 10GHz respectively for a fixed labelling error of 20\%. }
\label{fig:STFT_SNR}
\end{figure*}

Next, we consider the effect of SNR on the three algorithms in Fig.\ref{fig:STFT_SNR}. Again, there is significant improvement in the SSIM after denoising for all three algorithms across all three frequencies.
We observe that both the SparseDAE and the StackedSDAE are more robust to noise than DAE at low SNRs (from $-15dB$ to $5dB$) at 2.4GHz (Fig.\ref{fig:STFT_SNR}a) and 5GHz (Fig.\ref{fig:STFT_SNR}b). These results indicate that the use of sparsity constraints in the hidden layer representations result in increased robustness of the algorithm to Gaussian noise. Additional stacking layers also benefit for slightly higher SNR values. However, at very low SNR values ($<-10dB$), in Fig.\ref{fig:STFT_SNR}c corresponding to 10GHz radar frequency, additional stacking layers may be required to retain the robustness of the algorithm. The performance here is poorer due to weaker radar signals at 10GHz due to the gain offset.  
\subsection{High Resolution Range Profiles}
\begin{figure*}[ht]
\centering 
\subfigure[]{
\includegraphics[scale=0.30]{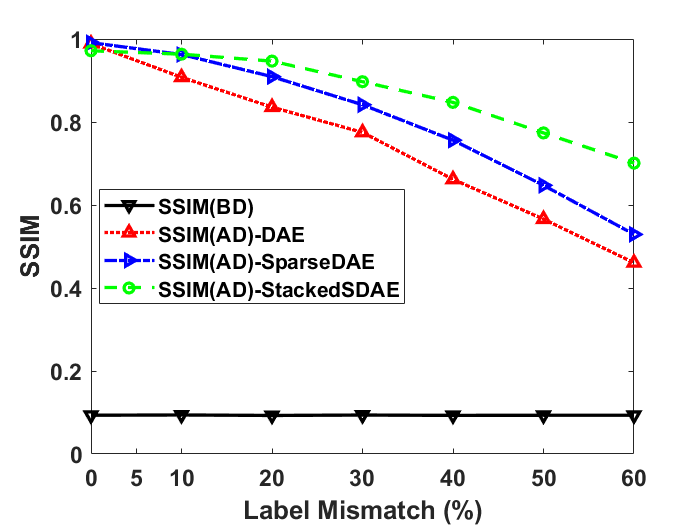}}
\subfigure[]{
\includegraphics[scale=0.30]{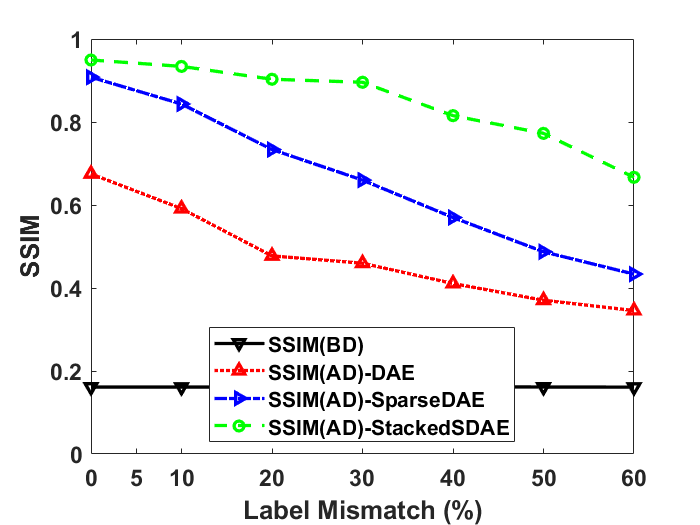}}
\subfigure[]{
\includegraphics[scale=0.30]{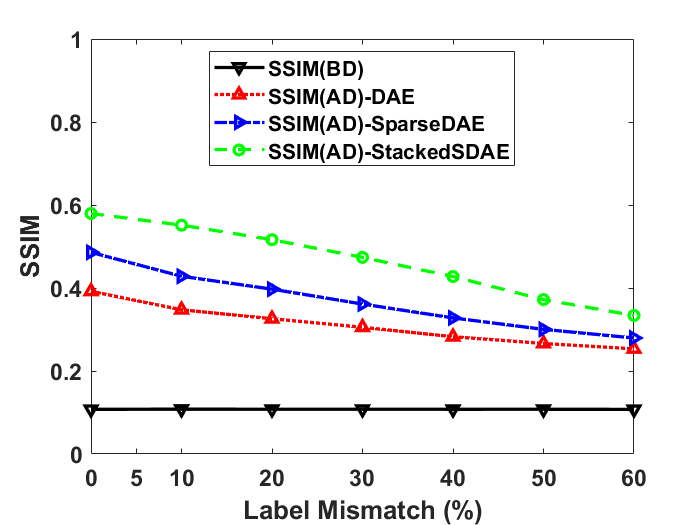}}
\caption{ SSIM variation for HRRPs with respect to label mismatch percentage at (a) 2.4GHz, (b) 5GHz, and (c) 10GHz respectively, for fixed SNR of $-10dB$. }
\label{fig:HRRP_labellingmismatch}
\end{figure*}

\begin{figure*}[ht]
\centering 
\subfigure[]{
\includegraphics[scale=0.30]{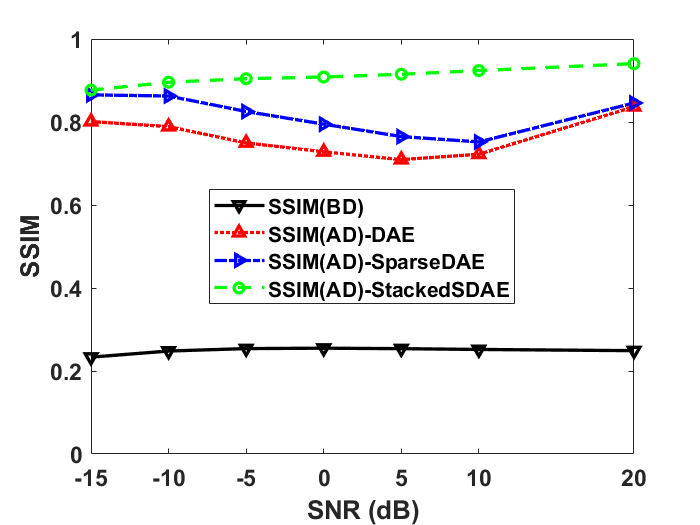}}
\subfigure[]{
\includegraphics[scale=0.30]{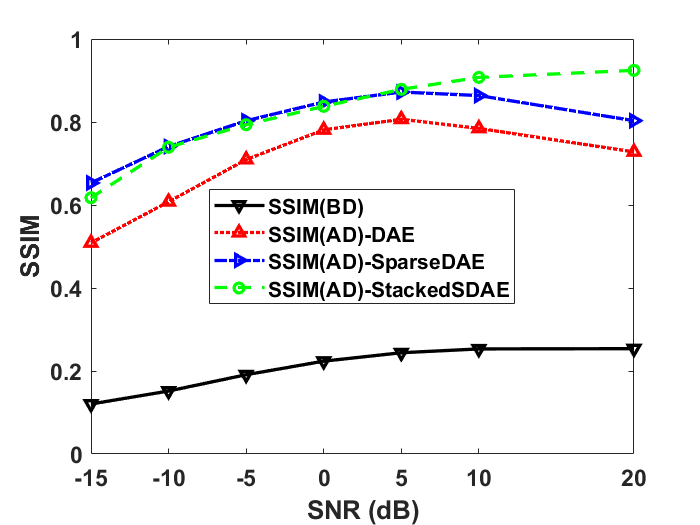}}
\subfigure[]{
\includegraphics[scale=0.30]{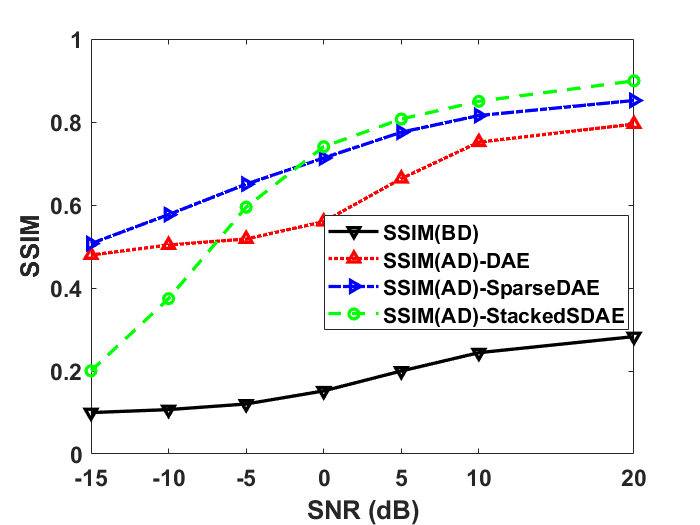}}
\caption{ SSIM variation for HRRPs with respect to SNR at (a) 2.4GHz, (b) 5GHz, and (c) 10GHz respectively for a fixed labelling error of 20\%. }
\label{fig:HRRP_results_SNR}
\end{figure*}

Next, we study the performance of the three algorithms on the HRRPs. First, we consider the label mismatch error in Fig.\ref{fig:HRRP_labellingmismatch} for the three carrier frequencies when SNR is $-10dB$. Again, we observe that all three algorithms result in increase in SSIM after denoising compared to the SSIM of the cluttered image before denoising. As the labelling mismatch error increases, the SSIM falls for all three algorithms across the three frequencies. However, the SSIM fall of the StackedSDAE is far less than DAE and SparseDAE and is at 0.75 even when the label mismatch is 50\% for 2.4GHz (Fig.\ref{fig:HRRP_labellingmismatch}a) and 5GHz (Fig.\ref{fig:HRRP_labellingmismatch}b). The denoising performances for all three algorithms are however significantly poorer at $10GHz$ possibly because of the greater multipath and weaker signal strength. 
Similar trends are observed when we compare the performance of the three algorithms for varying SNR in Fig.\ref{fig:HRRP_results_SNR}. 
Both SparseDAE and StackedDAE are more robust than DAE at low SNR values for 2.4GHz (Fig.\ref{fig:HRRP_results_SNR}a) and 5GHz (Fig.\ref{fig:HRRP_results_SNR}b). But StackedSDAE deteriorates significantly at extremely low SNR for 10GHz (Fig.\ref{fig:HRRP_results_SNR}c) below $-10dB$, due to the weaker signal strength, indicating that we may need greater depth in the hidden layers of the autoencoder. 
\section{Measurement Results}
\label{Sec:Measurements}
\subsection{Measurement Data Collection}
\label{Sec:Measurement_setup}
\begin{figure}[htbp] 
\centering
  \includegraphics[width=1.7in,height=1.5in]{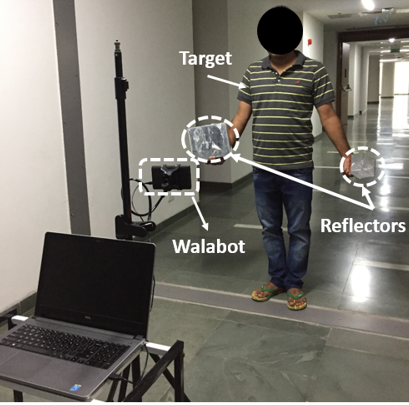}
  \hfil
  \caption{Measurement setup of frontal imaging of a slow moving human by Walabot }
  \label{fig:Exp_setup}
\end{figure}

In the previous sections, we demonstrated the effectiveness of the SparseDAE and StackedSDAE in denoising spectrograms and HRRPs generated in ACR scenarios.
However, these algorithms are essentially suited for denoising \emph{any} type of radar signature. To support this claim, we evaluate these algorithms on a third type of radar signature - the range enhanced frontal images. These images are generated by processing wideband measurement data from 3.3 to 10.3GHz captured using an imaging sensor called Walabot Pro \cite{walabot_2017}. This is an uncalibrated sensor that consists of a $4 \times 4$ antenna array with a maximum detectable range of about 4m in line-of-sight conditions and a field-of-view of approximately $90^{\circ}$ across azimuth and elevation. The radar data cube is processed through three-dimensional Fourier transform. Then the peaks across the range domain are superposed  to obtain range-enhanced frontal images of targets.  
The experimental setup for our measurement data collection is shown in Fig.\ref{fig:Exp_setup}. 
The DAE and its variants are trained with clean images of a slow moving human subject at 2m gathered in line-of-sight environments and the corresponding cluttered images. The  subject carries two boxes covered with aluminum tape to enhance the reflectivity from the hands. 
\begin{figure*}[ht] 
\centering
  \includegraphics[width=4in,height=3.1in]{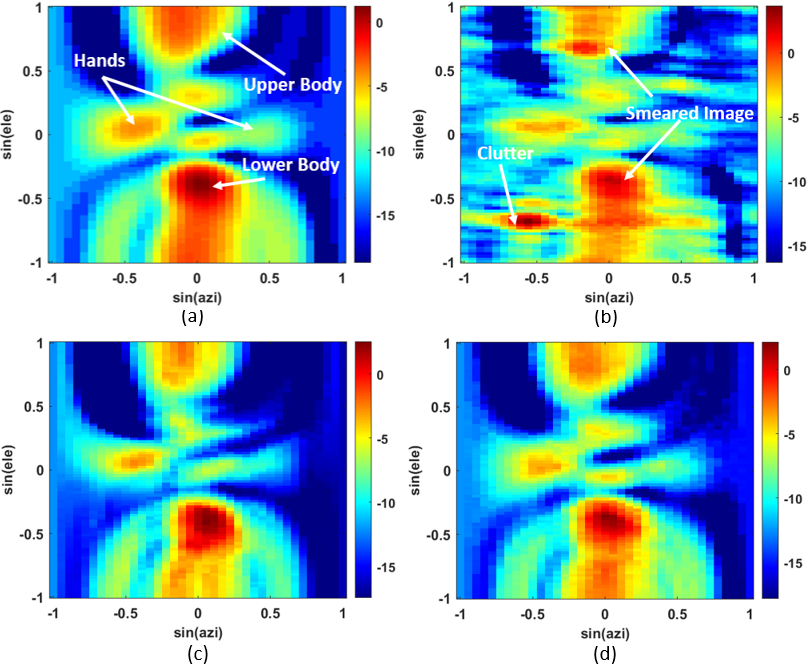}
  \caption{Measured range enhanced frontal images of a slow moving human (a) line-of-sight clean image, (b) cluttered Image with SCR=0dB, (c,d) reconstructed radar images after denoising for hidden layer dimensions $l_1$=$100$, and $l_2$=$500$ respectively.}
  \label{fig:measurement_images}
\end{figure*}
Fig. \ref{fig:measurement_images}(a) shows the frontal radar image of the subject in line-of-sight conditions where the torso, legs and arms of the human are noticeable. The experiments are performed on $5$ subjects of different heights and girth. For each of these subjects, we captured 90 measurements at different orientations with respect to the sensor.  

Due to the low transmitted power and limited dynamic range, the Walabot cannot be used in typical Indian through-wall scenarios (20cm brick walls). Instead, we synthetically corrupted the radar images with three types of distortions - additive Gaussian noise (of SNR from 0 to 30dB), clutter, and labelling errors. 
We modeled the clutter as a collection of discrete point scatterers randomly distributed across the radar's field-of-view whose magnitudes were varied to obtain signal to clutter ratios (SCR) spanning from 0 to 30dB. The phase of each scatterer followed a uniform distribution across $360^{\circ}$. Using Binomial distribution and a probability of false alarms of 0.06, we obtain approximately 5 false alarms for each image. We generated the cluttered images by the complex sum of the measurement, and clutter signals.  
An example of the distortions introduced to frontal image by clutter signals is shown in Fig. \ref{fig:measurement_images}(b) where we observe ghost targets. 

We introduced labelling mismatch errors by shuffling a proportion of the labels of the clean images so that they did not correspond correctly to the corrupted images. Each of the $31 \times 31$ pixel images were vectorized. 
Then the images are stacked column-wise to obtain a $[961 \times 450]$ matrix. The data is then split into a training set ($80\%$) and test set ($20\%$). 
\subsection{Results and Analyses}
\label{Sec:Measurement_results}
\begin{figure*}[ht]
\centering 
\subfigure[]{
\includegraphics[scale=0.3]{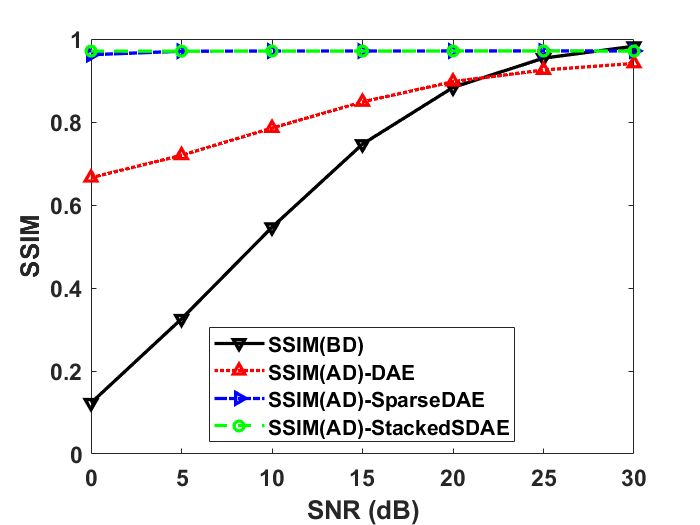}}
\subfigure[]{
\includegraphics[scale=0.3]{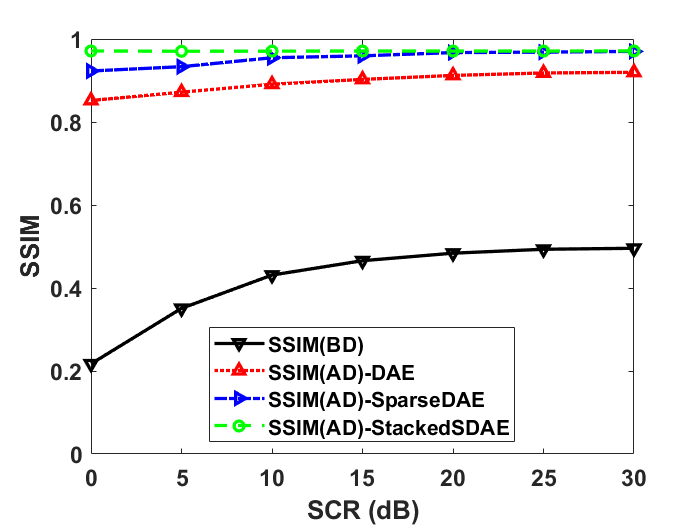}}\\
\subfigure[]{
\includegraphics[scale=0.3]{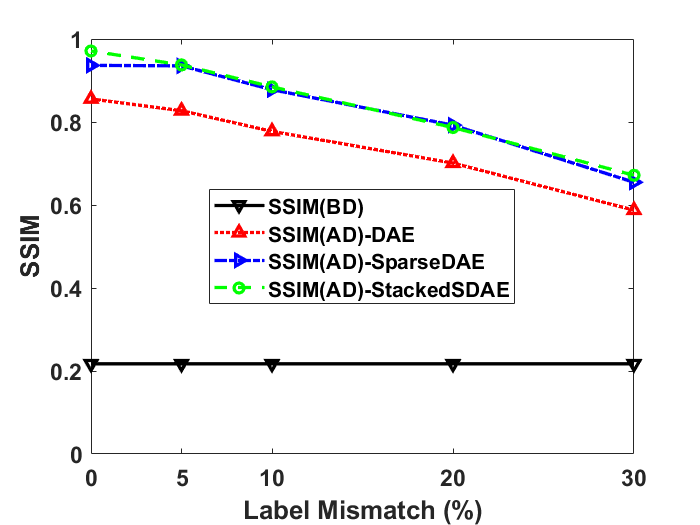}}
\subfigure[]{
\includegraphics[scale=0.3]{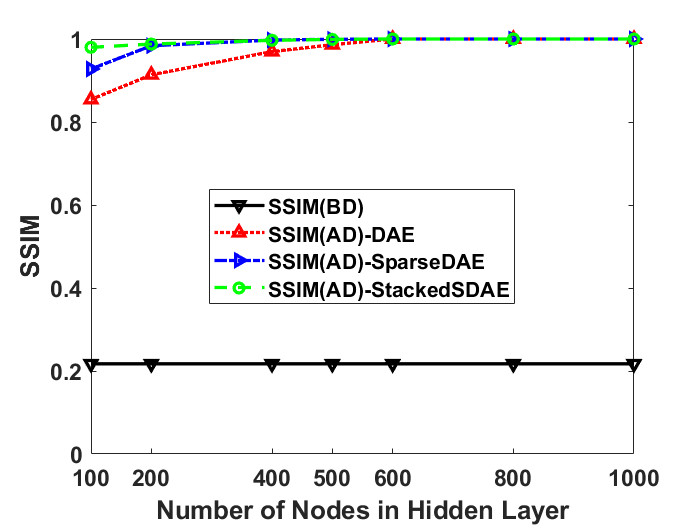}}
\caption{ SSIM variation with respect to  (a) { SNR, (b) signal to clutter ratio (SCR), (c) labelling mismatch error during training stage and (d) the number of nodes in the hidden layer }. }
\label{fig:var_num_nodes}
\end{figure*}

The training and testing of the algorithms is carried out in Matlab 2018b on Intel(R) Core(TM) i7-5500 processor with $16$-GB Ram running at $2.40$ GHz.
We present the results of the conventional DAE and its two variants - SparseDAE and StackedSDAE in Fig.\ref{fig:var_num_nodes}. In all the results, we present the SSIM between the corrupted and ground truth reference clean image \emph{before denoising (BD)} and the SSIM between the reconstructed image and clean image \emph{after denoising (AD)}. We observe a significant improvement in the SSIM after denoising for all three algorithms in all of the cases. 

First, we study the effect of SNR on the denoising performance in Fig.\ref{fig:var_num_nodes}(a). We observe that as the SNR decreases, the SSIM degrades significantly for conventional DAE. However, both SparseDAE and StackedSDAE are robust to noise since the performance does not significantly deteriorate with fall in SNR. 
The DAE, is however, less sensitive to low SCR values as observed in Fig.\ref{fig:var_num_nodes}b. The SparseDAE deteriorates slightly but the StackedSDAE does not get significantly impacted by low SCR values. 
The SSIM falls with increase in labelling error in Fig. \ref{fig:var_num_nodes}c. However, the SparseDAE and StackedSDAE are less sensitive than the conventional DAE. 
Finally, we examine the sensitivity of the algorithms' performance to the number of nodes in the hidden layer in Fig.\ref{fig:var_num_nodes}d. We observe that the performance of all three algorithms converge with an increase in the number of nodes. The StackedSDAE, however, converges for the fewest number of nodes. 
The algorithms require a minimum number of nodes in the hidden layers to achieve maximum improvement. The number of nodes is an important metric that determines the computational complexity (both time and memory) of the algorithm. 

We compare the computational time complexity of the three algorithms during training and test in Table.\ref{table:computational_complex}. 
\begin{table}[]
\centering
\caption{Computational time complexity of DAE and its variants}
\label{table:computational_complex}
\begin{tabular}{c c c}
\hline
\textbf{Algorithm} & \textbf{Training Time (sec)} & \textbf{Test Time (msec)} \\ \hline \hline
\begin{tabular}{@{}l@{}}\textbf{DAE}\end{tabular} & 205.3 & 76 \\ \hline
\begin{tabular}{@{}l@{}}\textbf{SparseDAE}\end{tabular} &  228.3 & 66\\ \hline
\begin{tabular}{@{}l@{}}\textbf{StackedSDAE}\end{tabular} & 466.3  & \textbf{0.7} \\ \hline
\end{tabular}
\end{table}
While the training time of DAE and SparseDAE are comparable, the StackedSDAE takes more than twice as long to train. This is because the incorporation of stacked hidden layers results in training requirements of additional weighting matrices in the hidden layers. However, due to the feature size reduction in these matrices, there is considerably lower computation time for the StackedSDAE during test. As a result, the StackedSDAE maybe more suitable for real time operations.  
\section{Benchmarking with other algorithms}
\begin{figure*}[ht]
    \centering
    \includegraphics[scale=0.65]{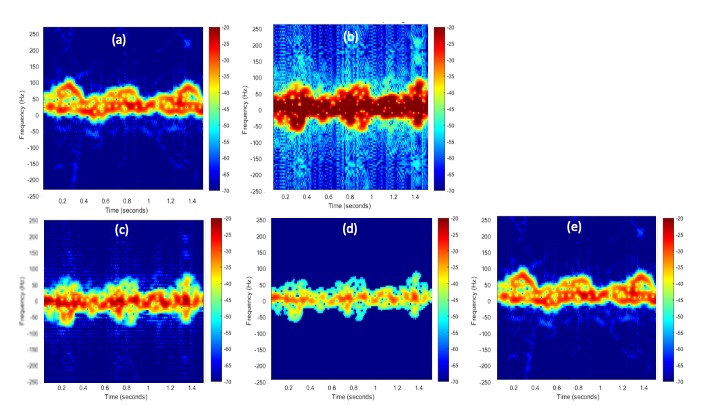}
    \caption{{Spectrograms of humans in (a) freespace and (b) ACR scenarios. Denoised images obtained using (c) wavelet filtering, (d) subspace filtering using SVD and (e) proposed methods using sparsity based autoencoders.}}
    \label{fig:ques2b}
\end{figure*}

We compare the performance of our algorithm, qualitatively and quantitatively, with subspace filtering based on singular value decomposition (SVD) and wavelet filtering.\\ 
\emph{Subspace Filtering:}
We have used subspace filtering methods presented in \cite{tivive2011wall} to denoise the simulated time-domain data used to generate the ACR spectrograms. Using SVD, we identify the eigen values of the data. The top few singular vectors belong to the signal while the remaining constitute the noise subspace. Then, we reconstruct denoised spectrograms after removing the distortions arising from the lower eigen values. Since the multipath clutter, in our scenario, are directly dependent on the target, they do not occupy orthogonal subspaces to the target subspace. The results presented in Fig.\ref{fig:ques2b} clearly indicate that the SVD based approach removes the noise but not the multipath distortions introduced in the ACR scenario, which are observed at negative Dopplers.

\emph{Wavelet Filtering:}
Next, we use wavelet based techniques for denoising the images. We apply discrete wavelet transform on the raw time-domain data. Again, we assume that the signal returns occupy the top wavelet coefficients while the remaining coefficients correspond to the noise and distortions. Therefore, we convert these coefficients to zero and then apply inverse discrete wavelet transform to reconstruct the denoised images. 
The resulting images in Fig.\ref{fig:ques2b} again show that the algorithm is successful in removing the independent noise but not the target dependent multipath distortions. In contrast, the proposed method based on autoencoders removes both the noise and the multipath clutter (negative Dopplers).

In Table.\ref{tab:ComparisonStateofArt}, we quantitatively compare the performances of the three algorithms. We observe that before denoising, the average SSIM between the noisy ACR signatures and the free space signatures are 0.05 and the normalized mean square error (NMSE) is 0.22. 
\begin{table*}[htbp]
    \centering
    \begin{tabular}{c|c|c|c|c}
    \hline
       {Method} & {SSIM (BD)} &{SSIM (AD)} & {NMSE (BD)} & {NMSE (AD)}  \\
        \hline
         {SVD} & {0.05} & {0.06}  & {0.22} & {0.1} \\
        {Wavelets} & {0.05} & {0.08} & {0.22} & {0.11} \\
         {StackedSDAE} &{0.05} & {0.92} & {0.22} & {0.01} \\
         \hline
    \end{tabular}
    \caption{Quantitative comparison of denoising performance by subspace filtering with SVD, wavelet filtering and proposed method using autoencoders.}
    \label{tab:ComparisonStateofArt}
\end{table*}
After denoising, the SSIM for SVD and wavelet filtering improve slightly by removing noise. However, the images still look different from the free space images due to the presence of the clutter. On the other hand, the proposed methods using autoencoders result in high SSIM (above 0.9) and a low NMSE (0.01) since they succeed in removing both noise and clutter based distortions.

\section{Conclusion}
Indoor radar signatures of dynamic human motions are corrupted by target dependent and independent static and dynamic clutter introduced by the presence of walls and other reflecting surfaces. We have used a variant of DAE, called the StackedSDAE, that incorporates both sparsity and depth in the hidden layer representations of the noisy images for clutter mitigation. The encoder and decoder stages of the algorithm are trained with labelled clean and noisy data. No additional information of the wall geometry or characteristics are required for the algorithm. Due to the additional stacked layers within the hidden layers, the training time for this algorithm is greater than that of the conventional DAE. However, each additional stacked layer has fewer nodes than the previous layer. This results in lowered feature size of the final representation and lower test time operation. The resulting denoised images are structurally similar to the radar images of the target that would have been obtained in free space scenarios. The StackedSDAE is more robust than the conventional DAE to labelling mismatch error between clean and noisy images during training. The algorithm is also more robust to low SNR ($-10$ to $+5dB$) than the conventional DAE. For example, in the case of the simulated ACR signatures, the denoised images have an SSIM above 0.75 even when the SNR is $-10dB$ and the label mismatch error is 50\%.
At extremely low SNR (below $-10dB$), we observe some deterioration in the performance possibly indicating that greater number of hidden layers are required at these SNRs. 
The complete database of radar signatures and algorithms are shared with the research community on the following URL \url{https://rb.gy/mmhzf6}.
\section*{Acknowledgement}
This work is sponsored by the 5IOA036 FA23861610004 grant by Air Force Office of Scientific Research (AFOSR), AOARD.
\bibliographystyle{ieeetran}
\bibliography{main}

\begin{thebibliography}{10}
\providecommand{\url}[1]{#1}
\csname url@samestyle\endcsname
\providecommand{\newblock}{\relax}
\providecommand{\bibinfo}[2]{#2}
\providecommand{\BIBentrySTDinterwordspacing}{\spaceskip=0pt\relax}
\providecommand{\BIBentryALTinterwordstretchfactor}{4}
\providecommand{\BIBentryALTinterwordspacing}{\spaceskip=\fontdimen2\font plus
\BIBentryALTinterwordstretchfactor\fontdimen3\font minus
  \fontdimen4\font\relax}
\providecommand{\BIBforeignlanguage}[2]{{%
\expandafter\ifx\csname l@#1\endcsname\relax
\typeout{** WARNING: IEEEtran.bst: No hyphenation pattern has been}%
\typeout{** loaded for the language `#1'. Using the pattern for}%
\typeout{** the default language instead.}%
\else
\language=\csname l@#1\endcsname
\fi
#2}}
\providecommand{\BIBdecl}{\relax}
\BIBdecl

\bibitem{amin2016through}
M.~G. Amin, \emph{Through-the-wall radar imaging}.\hskip 1em plus 0.5em minus
  0.4em\relax CRC press, 2016.

\bibitem{amin2017radar}
M.~Amin, \emph{Radar for indoor monitoring: Detection, classification, and
  assessment}.\hskip 1em plus 0.5em minus 0.4em\relax CRC Press, 2017.

\bibitem{gurbuz2019radar}
S.~Z. Gurbuz and M.~G. Amin, ``Radar-based human-motion recognition with deep
  learning: Promising applications for indoor monitoring,'' \emph{IEEE Signal
  Processing Magazine}, vol.~36, no.~4, pp. 16--28, 2019.

\bibitem{le2019radar}
J.~Le~Kernec, F.~Fioranelli, C.~Ding, H.~Zhao, L.~Sun, H.~Hong, J.~Lorandel,
  and O.~Romain, ``Radar signal processing for sensing in assisted living: The
  challenges associated with real-time implementation of emerging algorithms,''
  \emph{IEEE Signal Processing Magazine}, vol.~36, no.~4, pp. 29--41, 2019.

\bibitem{thayaparan2008micro}
T.~Thayaparan, L.~Stankovi{\'c}, and I.~Djurovi{\'c}, ``Micro-doppler-based
  target detection and feature extraction in indoor and outdoor environments,''
  \emph{Journal of the Franklin Institute}, vol. 345, no.~6, pp. 700--722,
  2008.

\bibitem{fioranelli2015classification}
F.~Fioranelli, M.~Ritchie, and H.~Griffiths, ``Classification of unarmed/armed
  personnel using the netrad multistatic radar for micro-doppler and singular
  value decomposition features,'' \emph{IEEE Geoscience and Remote Sensing
  Letters}, vol.~12, no.~9, pp. 1933--1937, 2015.

\bibitem{chen2003analysis}
V.~C. Chen, F.~Li, S.-S. Ho, and H.~Wechsler, ``Analysis of micro-doppler
  signatures,'' \emph{IEE Proceedings-Radar, Sonar and Navigation}, vol. 150,
  no.~4, pp. 271--276, 2003.

\bibitem{ram2008doppler}
S.~S. Ram, Y.~Li, A.~Lin, and H.~Ling, ``Doppler-based detection and tracking
  of humans in indoor environments,'' \emph{Journal of the Franklin Institute},
  vol. 345, no.~6, pp. 679--699, 2008.

\bibitem{narayanan2010through}
R.~M. Narayanan, M.~C. Shastry, P.-H. Chen, and M.~Levi, ``Through-the-wall
  detection of stationary human targets using doppler radar,'' \emph{Progress
  In Electromagnetics Research B}, vol.~20, pp. 147--166, 2010.

\bibitem{clemente2013developments}
C.~Clemente, A.~Balleri, K.~Woodbridge, and J.~J. Soraghan, ``Developments in
  target micro-doppler signatures analysis: radar imaging, ultrasound and
  through-the-wall radar,'' \emph{EURASIP Journal on Advances in Signal
  Processing}, vol. 2013, no.~1, p.~47, 2013.

\bibitem{ram2008through}
S.~S. Ram and H.~Ling, ``Through-wall tracking of human movers using joint
  doppler and array processing,'' \emph{IEEE Geoscience and Remote Sensing
  Letters}, vol.~5, no.~3, pp. 537--541, 2008.

\bibitem{lin2006frontal}
A.~Lin and H.~Ling, ``Frontal imaging of human using three-element doppler and
  direction-of-arrival radar,'' \emph{Electronics Letters}, vol.~42, no.~11,
  pp. 660--661, 2006.

\bibitem{ram2015high}
S.~S. Ram and A.~Majumdar, ``High-resolution radar imaging of moving humans
  using doppler processing and compressed sensing,'' \emph{IEEE Transactions on
  Aerospace and Electronic Systems}, vol.~51, no.~2, pp. 1279--1287, 2015.

\bibitem{ahmad2005synthetic}
F.~Ahmad, M.~G. Amin, and S.~A. Kassam, ``Synthetic aperture beamformer for
  imaging through a dielectric wall,'' \emph{IEEE transactions on aerospace and
  electronic systems}, vol.~41, no.~1, pp. 271--283, 2005.

\bibitem{dehmollaian2008refocusing}
M.~Dehmollaian and K.~Sarabandi, ``Refocusing through building walls using
  synthetic aperture radar,'' \emph{IEEE transactions on geoscience and remote
  sensing}, vol.~46, no.~6, pp. 1589--1599, 2008.

\bibitem{ahmad2008three}
F.~Ahmad, Y.~Zhang, and M.~G. Amin, ``Three-dimensional wideband beamforming
  for imaging through a single wall,'' \emph{IEEE Geoscience and remote sensing
  letters}, vol.~5, no.~2, pp. 176--179, 2008.

\bibitem{yoon2008high}
Y.-S. Yoon and M.~G. Amin, ``High-resolution through-the-wall radar imaging
  using beamspace music,'' \emph{IEEE Transactions on Antennas and
  Propagation}, vol.~56, no.~6, pp. 1763--1774, 2008.

\bibitem{cammenga2015high}
Z.~A. Cammenga, G.~E. Smith, and C.~J. Baker, ``High range resolution
  micro-doppler analysis,'' in \emph{SPIE Defense+ Security}.\hskip 1em plus
  0.5em minus 0.4em\relax International Society for Optics and Photonics, 2015,
  pp. 94\,611G--94\,611G.

\bibitem{martone2009through}
A.~Martone, K.~Ranney, and R.~Innocenti, ``Through-the-wall detection of
  slow-moving personnel,'' in \emph{Radar Sensor Technology XIII}, vol.
  7308.\hskip 1em plus 0.5em minus 0.4em\relax International Society for Optics
  and Photonics, 2009, p. 73080Q.

\bibitem{ahmad2013through}
F.~Ahmad and M.~G. Amin, ``Through-the-wall human motion indication using
  sparsity-driven change detection,'' \emph{IEEE Transactions on Geoscience and
  Remote Sensing}, vol.~51, no.~2, pp. 881--890, 2013.

\bibitem{gennarelli2015multiple}
G.~Gennarelli, G.~Vivone, P.~Braca, F.~Soldovieri, and M.~G. Amin, ``Multiple
  extended target tracking for through-wall radars,'' \emph{IEEE Transactions
  on Geoscience and Remote Sensing}, vol.~53, no.~12, pp. 6482--6494, 2015.

\bibitem{he2015range}
Y.~He, P.~Molchanov, T.~Sakamoto, P.~Aubry, F.~Le~Chevalier, and A.~Yarovoy,
  ``Range-doppler surface: a tool to analyse human target in ultra-wideband
  radar,'' \emph{IET Radar, Sonar \& Navigation}, vol.~9, no.~9, pp.
  1240--1250, 2015.

\bibitem{fogle2012micro}
O.~R. Fogle and B.~D. Rigling, ``Micro-range/micro-doppler decomposition of
  human radar signatures,'' \emph{IEEE Transactions on Aerospace and Electronic
  Systems}, vol.~48, no.~4, pp. 3058--3072, 2012.

\bibitem{erol2019radar}
B.~Erol and M.~G. Amin, ``Radar data cube processing for human activity
  recognition using multisubspace learning,'' \emph{IEEE Transactions on
  Aerospace and Electronic Systems}, vol.~55, no.~6, pp. 3617--3628, 2019.

\bibitem{zhao2018through}
M.~Zhao, T.~Li, M.~Abu~Alsheikh, Y.~Tian, H.~Zhao, A.~Torralba, and D.~Katabi,
  ``Through-wall human pose estimation using radio signals,'' in
  \emph{Proceedings of the IEEE Conference on Computer Vision and Pattern
  Recognition}, 2018, pp. 7356--7365.

\bibitem{sume2009radar}
A.~Sume, M.~Gustafsson, A.~Janis, S.~Nilsson, J.~Rahm, and A.~Orbom, ``Radar
  detection of moving objects around corners,'' in \emph{Radar Sensor
  Technology XIII}, vol. 7308.\hskip 1em plus 0.5em minus 0.4em\relax
  International Society for Optics and Photonics, 2009, p. 73080V.

\bibitem{sume2011radar}
A.~Sume, M.~Gustafsson, M.~Herberthson, A.~Janis, S.~Nilsson, J.~Rahm, and
  A.~Orbom, ``Radar detection of moving targets behind corners,'' \emph{IEEE
  Transactions on Geoscience and Remote Sensing}, vol.~49, no.~6, pp.
  2259--2267, 2011.

\bibitem{rabaste2015around}
O.~Rabaste, E.~Colin-Koeniguer, D.~Poullin, A.~Cheraly, J.-F. Petex, and H.-K.
  Phan, ``Around-the-corner radar: detection of a human being in non-line of
  sight,'' \emph{IET Radar, Sonar \& Navigation}, vol.~9, no.~6, pp. 660--668,
  2015.

\bibitem{rabaste2017around}
K.-P.-H. Thai, O.~Rabaste, J.~Bosse, D.~Poullin, I.~Hinostroza, T.~Letertre,
  T.~Chonavel \emph{et~al.}, ``Around-the-corner radar: Detection and
  localization of a target in non-line of sight,'' in \emph{2017 IEEE Radar
  Conference (RadarConf)}.\hskip 1em plus 0.5em minus 0.4em\relax IEEE, 2017,
  pp. 0842--0847.

\bibitem{vishwakarma2020micro}
S.~Vishwakarma, A.~Rafiq, and S.~S. Ram, ``Micro-doppler signatures of dynamic
  humans from around the corner radar,'' in \emph{2020 IEEE International Radar
  Conference (RADAR)}.\hskip 1em plus 0.5em minus 0.4em\relax IEEE, 2020, pp.
  169--174.

\bibitem{ahmad2007autofocusing}
F.~Ahmad, M.~G. Amin, and G.~Mandapati, ``Autofocusing of through-the-wall
  radar imagery under unknown wall characteristics,'' \emph{IEEE transactions
  on image processing}, vol.~16, no.~7, pp. 1785--1795, 2007.

\bibitem{leigsnering2014multipath}
M.~Leigsnering, F.~Ahmad, M.~Amin, and A.~Zoubir, ``Multipath exploitation in
  through-the-wall radar imaging using sparse reconstruction,'' \emph{IEEE
  Transactions on Aerospace and Electronic Systems}, vol.~50, no.~2, pp.
  920--939, 2014.

\bibitem{vishwakarma2018mitigation}
S.~Vishwakarma, V.~Ummalaneni, M.~S. Iqbal, A.~Majumdar, and S.~S. Ram,
  ``Mitigation of through-wall interference in radar images using denoising
  autoencoders,'' in \emph{IEEE Radar Conference (RadarConf18)}.\hskip 1em plus
  0.5em minus 0.4em\relax IEEE, 2018, pp. 1543--1548.

\bibitem{vishwakarma2020mitigation}
S.~Vishwakarma and S.~S. Ram, ``Mitigation of through-wall distortions of
  frontal radar images using denoising autoencoders,'' \emph{IEEE Transactions
  on Geoscience and Remote Sensing}, 2020.

\bibitem{chen2016multipath}
X.~Chen and W.~Chen, ``Multipath ghost elimination for through-wall radar
  imaging,'' \emph{IET Radar, Sonar \& Navigation}, vol.~10, no.~2, pp.
  299--310, 2016.

\bibitem{tan2014multipath}
Q.~Tan, H.~Leung, Y.~Song, and T.~Wang, ``Multipath ghost suppression for
  through-the-wall radar,'' \emph{IEEE Transactions on Aerospace and Electronic
  Systems}, vol.~50, no.~3, pp. 2284--2292, 2014.

\bibitem{leigsnering2018sparsity}
M.~Leigsnering, \emph{Sparsity-Based Multipath Exploitation for
  Through-the-Wall Radar Imaging}.\hskip 1em plus 0.5em minus 0.4em\relax
  Springer, 2018.

\bibitem{tang2017multipolarization}
V.~H. Tang, A.~Bouzerdoum, and S.~L. Phung, ``Multipolarization through-wall
  radar imaging using low-rank and jointly-sparse representations,'' \emph{IEEE
  Transactions on Image Processing}, vol.~27, no.~4, pp. 1763--1776, 2017.

\bibitem{vishwakarma2017detection}
S.~Vishwakarma and S.~S. Ram, ``Detection of multiple movers based on single
  channel source separation of their micro-dopplers,'' \emph{IEEE Transactions
  on Aerospace and Electronic Systems}, vol.~54, no.~1, pp. 159--169, 2017.

\bibitem{DAE}
\BIBentryALTinterwordspacing
P.~Vincent, H.~Larochelle, I.~Lajoie, Y.~Bengio, and P.-A. Manzagol, ``Stacked
  denoising autoencoders: Learning useful representations in a deep network
  with a local denoising criterion,'' \emph{J. Mach. Learn. Res.}, vol.~11, pp.
  3371--3408, Dec. 2010. [Online]. Available:
  \url{http://dl.acm.org/citation.cfm?id=1756006.1953039}
\BIBentrySTDinterwordspacing

\bibitem{vincent2008extracting}
P.~Vincent, H.~Larochelle, Y.~Bengio, and P.-A. Manzagol, ``Extracting and
  composing robust features with denoising autoencoders,'' in \emph{Proceedings
  of the 25th international conference on Machine learning}.\hskip 1em plus
  0.5em minus 0.4em\relax ACM, 2008, pp. 1096--1103.

\bibitem{hinton2006reducing}
G.~E. Hinton and R.~R. Salakhutdinov, ``Reducing the dimensionality of data
  with neural networks,'' \emph{science}, vol. 313, no. 5786, pp. 504--507,
  2006.

\bibitem{sakurada2014anomaly}
M.~Sakurada and T.~Yairi, ``Anomaly detection using autoencoders with nonlinear
  dimensionality reduction,'' in \emph{Proceedings of the MLSDA 2014 2nd
  Workshop on Machine Learning for Sensory Data Analysis}.\hskip 1em plus 0.5em
  minus 0.4em\relax ACM, 2014, p.~4.

\bibitem{socher2011dynamic}
R.~Socher, E.~H. Huang, J.~Pennin, C.~D. Manning, and A.~Y. Ng, ``Dynamic
  pooling and unfolding recursive autoencoders for paraphrase detection,'' in
  \emph{Advances in neural information processing systems}, 2011, pp. 801--809.

\bibitem{chen2012marginalized}
M.~Chen, Z.~Xu, K.~Weinberger, and F.~Sha, ``Marginalized denoising
  autoencoders for domain adaptation,'' \emph{arXiv preprint arXiv:1206.4683},
  2012.

\bibitem{ng2011sparse}
A.~Ng \emph{et~al.}, ``Sparse autoencoder,'' \emph{CS294A Lecture notes},
  vol.~72, no. 2011, pp. 1--19, 2011.

\bibitem{vishwakarma2019Clu}
S.~Vishwakarma, N.~Pandey, and S.~S. Ram, ``Clutter mitigation in range
  enhanced radar images using sparsity based denoising autoencoders,'' in
  \emph{2019 International Radar Conference (RADAR)}.\hskip 1em plus 0.5em
  minus 0.4em\relax IEEE, 2019, pp. 1--6.

\bibitem{walabot_2017}
``Walabot,''
  \url{https://cdn.sparkfun.com/assets/learn_tutorials/7/2/4/walabot-tech-brief-416.pdf/},
  2017 (accessed January 18, 2020), [Online].

\bibitem{mehta2016stacked}
J.~Mehta, K.~Gupta, A.~Gogna, A.~Majumdar, and S.~Anand, ``Stacked robust
  autoencoder for classification,'' in \emph{International Conference on Neural
  Information Processing}.\hskip 1em plus 0.5em minus 0.4em\relax Springer,
  2016, pp. 600--607.

\bibitem{MAJUMDAR2018271}
A.~Majumdar, ``Graph structured autoencoder,'' \emph{Neural Networks}, vol.
  106, pp. 271 -- 280, 2018.

\bibitem{boyd2011distributed}
S.~Boyd, N.~Parikh, E.~Chu, B.~Peleato, and J.~Eckstein, ``Distributed
  optimization and statistical learning via the alternating direction method of
  multipliers,'' \emph{Foundations and Trends{\textregistered} in Machine
  Learning}, vol.~3, no.~1, pp. 1--122, 2011.

\bibitem{bishop2006pattern}
C.~M. Bishop, ``Pattern recognition,'' \emph{Machine Learning}, vol. 128, pp.
  1--58, 2006.

\bibitem{wang2004image}
Z.~Wang, A.~C. Bovik, H.~R. Sheikh, and E.~P. Simoncelli, ``Image quality
  assessment: from error visibility to structural similarity,'' \emph{IEEE
  transactions on image processing}, vol.~13, no.~4, pp. 600--612, 2004.

\bibitem{smith2012stochastic}
S.~M. Smith and C.~Furse, ``Stochastic fdtd for analysis of statistical
  variation in electromagnetic fields,'' \emph{IEEE Transactions on Antennas
  and Propagation}, vol.~60, no.~7, pp. 3343--3350, 2012.

\bibitem{ram2008simulation}
S.~S. Ram and H.~Ling, ``Simulation of human microdopplers using computer
  animation data,'' in \emph{2008 IEEE Radar Conference}.\hskip 1em plus 0.5em
  minus 0.4em\relax IEEE, 2008, pp. 1--6.

\bibitem{ram2010simulation}
S.~S. Ram, C.~Christianson, Y.~Kim, and H.~Ling, ``Simulation and analysis of
  human micro-dopplers in through-wall environments,'' \emph{IEEE Transactions
  on Geoscience and remote sensing}, vol.~48, no.~4, pp. 2015--2023, 2010.

\bibitem{tivive2011wall}
F.~H.~C. Tivive, M.~G. Amin, and A.~Bouzerdoum, ``Wall clutter mitigation based
  on eigen-analysis in through-the-wall radar imaging,'' in \emph{Digital
  Signal Processing (DSP), 2011 17th International Conference on}.\hskip 1em
  plus 0.5em minus 0.4em\relax IEEE, 2011, pp. 1--8.

\end{thebibliography}
\end{document}